\documentclass{nature}
\usepackage{aas_macros}
\usepackage{amssymb}

\usepackage{color}

\usepackage{hyperref} 

\def    \apjl  		{\it {Astrophys.~J.~Lett.}}
\def    \apj  		{\it {Astrophys.~J.}}
\def    \aj  		{\it {Astron.~J.}}
\def    \apjs  		{\it {Astrophy.~J.~Suppl.}}

\def    \mnras  	{\it {Mon.~Not.~R.~Astron.~Soc.}}
\def    \araa  		{\it {Annu.~Rev.~Astron.~Astrophys.}}
\def    \aa       {\it {Astron.~Astrophys.}}

\title{Rotational Disruption of Dust Grains by Radiative Torques in Strong Radiation Fields}

\author{Thiem Hoang$^{\ref{TH1},\ref{TH2}}$, Le Ngoc Tram$^{\ref{TH1},\ref{Tr2},\ref{Tr3}}$, Hyeseung Lee$^{\ref{TH1}}$, and Sang-Hyeon Ahn$^{\ref{TH1}}$}

\begin{document}

\maketitle

\begin{affiliations}
 \item 
\label{TH1}Korea Astronomy and Space Science Institute, Daejeon 34055, Republic of Korea
\item
\label{TH2} Korea University of Science and Technology, 217 Gajeong-ro, Yuseong-gu, Daejeon, 34113, Republic of Korea
\item
\label{Tr2} University of Science and Technology of Hanoi, VAST, 18 Hoang Quoc Viet, Vietnam
\item
\label{Tr3} Current address: SOFIA-USRA, NASA Ames Research Center, MS 232-11, Moffett Field, CA 94035, USA
\end{affiliations}

\begin{abstract}
Massive stars, supernovae, and kilonovae are among the most luminous radiation sources in the universe. Observations usually show near- to mid-infrared (NIR--MIR, $\lambda\sim 1-5~\mu$m) emission excess from H\,{\sc ii} regions around young massive star clusters (YMSCs). Early phase observations in optical to NIR wavelengths of type Ia supernovae also reveal unusual properties of dust extinction and dust polarization. The popular explanation for such NIR-MIR excess and unusual dust properties is the predominance of small grains (size $a\lesssim 0.05~\mu$m) relative to large grains ($a\gtrsim 0.1~\mu$m) in the local environment of these strong radiation sources. The question of why small grains are predominant in these environments remains a mystery. Here we report a new mechanism of dust destruction based on centrifugal stress within extremely fast-rotating grains spun-up by radiative torques, which we term the RAdiative Torque Disruption (RATD) mechanism. We find that RATD can disrupt large grains located within a distance of $\sim 1$ pc from a massive star of luminosity $L\sim 10^{4}L_{\odot}$ or a supernova. This effect increases the abundance of small grains relative to large grains and successfully reproduces the observed NIR-MIR excess and anomalous dust extinction/polarization. We apply the RATD mechanism for kilonovae and find that dust within $\sim$ 0.1 pc would be dominated by small grains. Small grains produced by RATD can also explain the steep far-UV rise in extinction curves toward starburst and high redshift galaxies, and the decrease of the escape fraction of Ly$\alpha$ photons from H\,{\sc ii} regions surrounding YMSCs.

\end{abstract}
\newpage
\section*{\large Introduction}\label{sec:intro}
Massive (OB) stars are the most luminous stellar objects in the universe, which can produce high luminosities of $L\sim 10^{4}-10^{6}L_{\odot}$ over millions of years. Such a strong radiation source can have a significant impact on the physical and chemical properties of gas and dust in the surrounding environment. Observations usually show a near- to mid- infrared (NIR--MIR, $\lambda\sim 1-5~\mu$m) emission excess from H\,{\sc ii} regions surrounding young massive star clusters (YMSCs)\cite{2008AJ....136.1415R,2013A&A...552A.140R}. The NIR-MIR excess is explained by the presence of a large fraction of small grains ($a\lesssim 0.05~\mu$m) that are transiently heated to high temperatures ($\sim 1,000$ K) by UV photons (see ref. \cite{2017ApJ...843...95M} and references therein). Such an explanation is difficult to reconcile with the fact that thermal sputtering is efficient in H\,{\sc ii} regions\cite{1979ApJ...231...77D} which can destroy small grains on a timescale of $\sim 10^{4}$ yr, much shorter than the age of YMSCs. Therefore, the remaining question is how small grains can be rapidly replenished against thermal sputtering in H\,{\sc ii} regions.

Type Ia supernovae (SNe Ia) have widely been used as standard candles to measure the expansion of the universe due to their stable intrinsic luminosity\cite{1998AJ....116.1009R}. To achieve the most precise constraints on cosmological parameters, the effect of dust extinction on the instrinsic light curve of SNe Ia must be accurately characterized. Optical to near-infrared photometric observations of SNe Ia during the early phase (i.e., within a few weeks after maximum brightness) reveal unusual properties of dust extinction, with unprecedented low values of the total-to-selective extinction ratio of $R_{\rm V}\lesssim 2$ (refs \cite{2008A&A...487...19N,2014ApJ...789...32B}), much lower than the standard Milky Way value of $R_{\rm V}\sim 3.1$ (ref. \cite{2003ARA&A..41..241D}). Moreover, polarimetric observations also report unusually low wavelengths of the maximum polarization ($\lambda_{\rm max}<0.4~\mu$m) for several SNe Ia\cite{Kawabata:2014gy,Patat:2015bb}. Numerical modeling of dust extinction\cite{2016P&SS..133...36N} and polarization curves\cite{2017ApJ...836...13H} toward individual SNe Ia demonstrate that the anomalous values of $R_{\rm V}$ and $\lambda_{\rm max}$ can be reproduced by the enhancement in the relative abundance of small grains to large grains in the host galaxy. 

Type II supernovae (SNe II) are considered the major source of dust formation\cite{2010A&A...518L.138B,2011Sci...333.1258M,2012ApJ...760...96G,Chawner:2018dn} in the early universe where dust formation by AGB stars is not significant\cite{2003MNRAS.343..427M}. Observational and theoretical studies show that grains in the early universe essentially have small sizes of $a\lesssim 0.05~\mu$m\cite{2014MNRAS.439.3073Y}. Moreover, early-phase observations of individual SNe II-P also reveal anomalous values of $R_{V}\sim 1.4-1.5$  (refs \cite{2009ApJ...694.1067P,2010ApJ...715..833O}), which requires the excess of small grains to resolve. Here we should distinguish {\it original} dust probed by early-phase observations of SNe with {\it new} dust formed in the supernova ejecta observed at later stages that might have large grains\cite{2013ApJ...774....8T,2015ApJ...801..141O,Gall:2014dk}. 

The previously known mechanisms of dust destruction\cite{1979ApJ...231..438D} cannot explain why small grains are predominant along the lines of sight towards SNe Ia and II-P, as revealed by early phase observations. 
First, thermal sublimation, which is efficient in strong radiation fields of SNe, would reduce rather than enhance the abundance of small grains\cite{1989ApJ...345..230G,2015ApJ...806..255H}. Second, shattering by grain-grain collisions in supernova shocks occurs at a late stage such that it cannot explain why early phase observations already show the predominance of small grains\cite{Patat:2015bb,2017ApJ...836...13H}. As we will see in  Methods (Section `Comparison of RATD to other destruction mechanisms'), a new mechanism based on grain-grain collisions induced by supernova radiation pressure\cite{2017ApJ...836...13H} cannot reproduce an excess of small grains on a timescale of several months observed toward SNe Ia and SNe II-P.

In this paper, we report a new mechanism of dust destruction based on centrifugal stress within extremely fast rotating grains spun-up by radiative torques, so-called RAdiative Torque Disruption (RATD) mechanism, which can successfully explain the aforementioned observational puzzles. We will also apply the RATD mechanism to study the feedback of kilonovae\cite{1998ApJ...507L..59L,2017PASJ...69..102T} on dust properties in the surrounding environments. An accurate description of dust extinction to kilonovae (KNe) is needed to infer intrinsic colour and light curves, as well as a precise measurement of opacity caused by lanthanides in the kilonova ejecta\cite{2017ApJ...849L..19G}.
\newpage

\section*{\large The dust destruction mechanism}\label{sec:result}
\subsection{Rotational Disruption by Radiative Torques.}
A dust grain of irregular shape subject to an anisotropic radiation field experiences radiative torques (RATs) due to differential scattering and absorption of incident photons\cite{1976Ap&SS..43..291D,1996ApJ...470..551D,2007MNRAS.378..910L}. 

For an average diffuse interstellar radiation field (ISRF) in the solar neighborhood\cite{1983A&A...128..212M}, irregular grains of effective size $a\sim 0.1~\mu$m can be spun-up to angular velocity of $\omega_{\rm RAT}\sim 10^{6}~\rm rad~s^{-1}$ by RATs\cite{1996ApJ...470..551D,2009ApJ...695.1457H}. In strong radiation fields such as near a massive star or a supernova where the radiation energy density can be increased by a factor $U\gg 1$ from the average ISRF, grains can be spun-up to extremely fast rotation with
\begin{eqnarray}
\omega_{\rm RAT}&\simeq &7.6\times 10^{10}\gamma a_{-5}^{1.7}\bar{\lambda}_{0.5}^{-1.7}U_{6}^{1/3}~\rm rad~s^{-1}~\label{eq:omega_RAT}
\end{eqnarray}
for $a\lesssim \bar{\lambda}/1.8$, and
\begin{eqnarray}
\omega_{\rm RAT}&\simeq &1.1\times 10^{12}a_{-5}^{-1}\gamma \bar{\lambda}_
{0.5}^{-1.7}U_{6}^{1/3} ~\rm rad~s^{-1}~~~\label{eq:omega_RAT1}
\end{eqnarray}
for $a> \bar{\lambda}/1.8$. Here $a_{-5}=a/(10^{-5}~\rm cm)$, $\gamma$ is the anisotropy degree of radiation, $\bar{\lambda}$ is the mean wavelength of the radiation field with $\bar{\lambda}_{0.5}=\bar{\lambda}/(0.5~\mu\rm m)$, and $U_{6}=U/10^{6}$ (see Methods for details). 

Due to a centrifugal force, a spinning grain develops a centrifugal stress that tends to tear the grain apart. The centrifugal stress averaged over the surface parallel to the spinning axis is given by $S=\rho\omega^{2}a^{2}/4$, with $\rho$ being the grain mass density (see Methods section `Centrifugal stress within a spinning grain'). When the centrifugal stress exceeds the maximum tensile strength of the grain material, $S_{\rm max}$, the grain is instantaneously disrupted into small fragments. We term this mechanism as RAdiative Torque Disruption (RATD).

The critical rotation rate above which the grain is disrupted can be obtained by equating the centrifugal stress $S$ to $S_{\rm max}$, which yields
\begin{eqnarray}
\omega_{\rm disr}=\frac{2}{a}\left(\frac{S_{\rm max}}{\rho} \right)^{1/2}\simeq 3.6\times 10^{9}a_{-5}^{-1}\hat{\rho}^{-1/2}S_{\rm max,9}^{1/2}~\rm rad~s^{-1},\label{eq:omega_cri}
\end{eqnarray}
where $\hat{\rho}=\rho/(3~\rm g~cm^{-3})$ and $S_{\rm max,9}=S_{\rm max}/(10^{9}~\rm erg~ cm^{-3})$. 

The critical rotation rate depends on the grain tensile strength, which is uncertain for interstellar dust. Here, we take $S_{\rm max}=10^{9}~\rm erg~cm^{-3}$ as a typical value for compact grains and $S_{\rm max}=10^{7}~\rm erg~cm^{-3}$ for composite grains (see Supplementary Information section `Tensile strengths of grain materials'). 

Due to the rapid increase of $\omega_{\rm RAT}$ and decrease of $\omega_{\rm disr}$ with the grain size $a$ (equations (\ref{eq:omega_RAT}) and (\ref{eq:omega_cri})), we expect large grains to be destroyed by RATD, whereas small grains can survive in strong radiation fields. In the following, we will calculate the critical size of grain disruption by RATs and its required timescale in the strong radiation fields of massive stars, supernovae, and kilonovae.

\subsection{Grain Disruption in Massive Stars and Young Massive Star Clusters.}\label{sec:MS}

For a given radiation field of constant bolometric luminosity $L$ and mean wavelength $\bar{\lambda}$, one can calculate $\omega_{\rm RAT}$ for a grid of grain sizes, assuming the gas density ($n_{\rm H}$) and temperature ($T_{\rm gas}$) for the local environment. The disruption size, $a_{\rm disr}$, is then obtained by comparing $\omega_{\rm RAT}$ with $\omega_{\rm disr}$. The time required to spin-up dust grains to $\omega_{\rm disr}$ (i.e., disruption time, $t_{\rm disr}$) is calculated by equation (\ref{eq:tdisr}) (see Methods for details). We calculate $a_{\rm disr}$ and $t_{\rm disr}$ for the physical parameters of the standard interstellar medium (ISM) with $n_{\rm H}=30~\rm cm^{-3}$ and $T_{\rm gas}=100~\rm K$, and an H\,{\sc ii} region with $n_{\rm H}=1.0~\rm cm^{-3}$ and $T_{\rm gas}= 10^{6}~\rm K$.

\begin{figure*}
\centering
\includegraphics[scale=0.45]{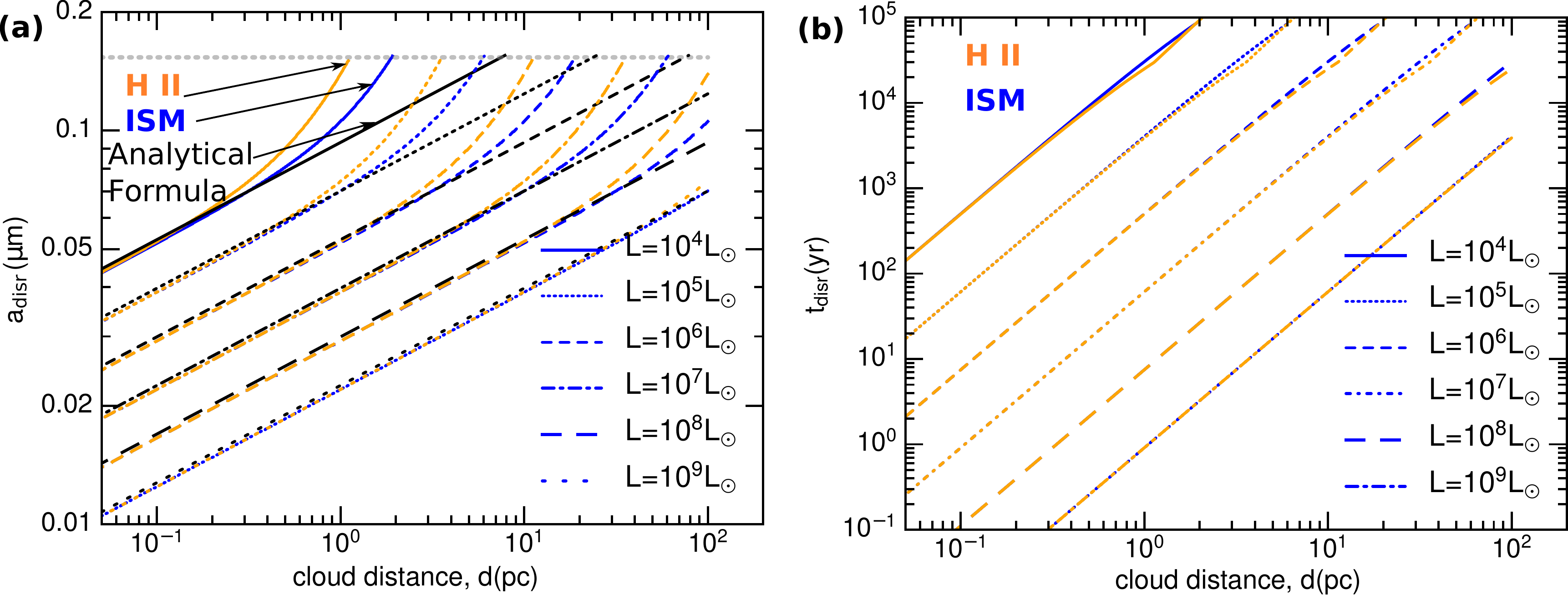}
\caption{{\bf Grain disruption size and disruption time vs. cloud distance from the central source for massive stars and YMSCs of different luminosity, assuming grain tensile strength $S_{\rm max}=10^{9}~\rm erg~cm^{-3}$}. Panel ({\bf a}): grain disruption size vs. cloud distance computed for the ISM (blue lines) and H\,{\sc ii} regions (orange lines). Results obtained from an analytical formula in the absence of gas damping (equation \ref{eq:a_cri} (Methods)) are shown in black lines. The horizontal line in the top marks $a_{\rm disr}=\bar{\lambda}/1.8$. Panel ({\bf b}): grain disruption time vs. cloud distance computed for the ISM and H\,{\sc ii} regions. The disruption time is short, below  $\sim 1$ Myr for YMSCs of $L\sim 10^{6}-10^{9}L_{\odot}$.}
\label{fig:atdisr_dis}
\end{figure*}

Figure \ref{fig:atdisr_dis} (panel (a)) shows the grain disruption size as a function of the cloud distance for $L= 10^{4}-10^{9}L_{\odot}$ for the ISM (blue lines) and H\,{\sc ii} regions (orange lines). The results obtained from an analytical formula (equation (\ref{eq:a_cri}) in Methods) where the grain rotational damping by gas collisions is disregarded (see Supplementary Information section `Grain rotational damping') is shown in black lines for comparison. 

The disruption size $a_{\rm disr}$ increases rapidly with increasing cloud distance and reaches $a_{\rm disr}\sim \bar{\lambda}/1.8\sim 0.16~\mu$m (marked by a horizontal line in the figure) at some distance. Beyond this distance, grain disruption ceases to occur due to the decrease of radiation energy density (see Fig. \ref{fig:atdisr_dis}). For $L= 10^{4}L_{\odot}$, which is typical for OB stars, we get $a_{\rm disr}\sim 0.1~\mu$m for $d\sim 1\rm pc$ for the ISM. For more luminous stars of $L= 10^{6}L_{\odot}$, $a_{\rm disr}\sim 0.1~\mu$m for $d\sim 10$ pc (see dashed line). For a YMSC of $L=10^{9}L_{\odot}$, one obtains $a_{\rm disrp}\sim 0.05~\mu$m for $d\sim 30$ pc, and $a_{\rm disr}\sim 0.02~\mu$m for $d\sim 1$ pc (see also Table \ref{tab:adisr_ISM}). 

For a given $L$, $a_{\rm disr}$ for the ISM and H\,{\sc ii} regions is similar at small distances. At large distances from the source, $a_{\rm disr}$ for H\,{\sc ii} regions is larger than for the ISM and for the case without gas damping (black lines in Fig. 1). The reason is that at large distances, rotational damping by gas collisions becomes dominant over the rotational damping by infrared emission, resulting in the increase of $a_{\rm disr}$ with the gas damping rate which scales as $n_{\rm H}T_{\rm gas}^{1/2}$ (Methods equation (\ref{eq:taudamp_gas})). This can also be seen through the increase in the critical radiation strength required to disrupt grains with the gas damping rate (see Methods and  Supplementary Figure 1). 

Figure \ref{fig:atdisr_dis} (panel (b)) shows the disruption time $t_{\rm disr}$ of $a=a_{\rm disr}$ grains as a function of the cloud distance for the different values of $L$. The disruption time increases rapidly with the cloud distance and decreases with increasing $L$. For grains at 10 pc, one obtains $t_{\rm disr}\sim$ 50--30,000 yr for $L \sim 10^{9}-10^{6}L_{\odot}$. This disruption time is much shorter than the age of YMSCs (a few Myr old).

\subsection{Grain Disruption in Supernovae and Kilonovae.}\label{sec:SNe_KN}

For time-varying radiation sources such as SNe and KNe, we first solve equation (\ref{eq:domega}) to obtain the time-dependent angular velocity $\omega(t)$. We then compare $\omega(t)$ with $\omega_{\rm disr}$ (equation (\ref{eq:omega_cri})) to determine grain disruption size and disruption time (see more details in Methods and Supplementary Figure 2).

\begin{figure*}
\centering
\includegraphics[scale=0.45]{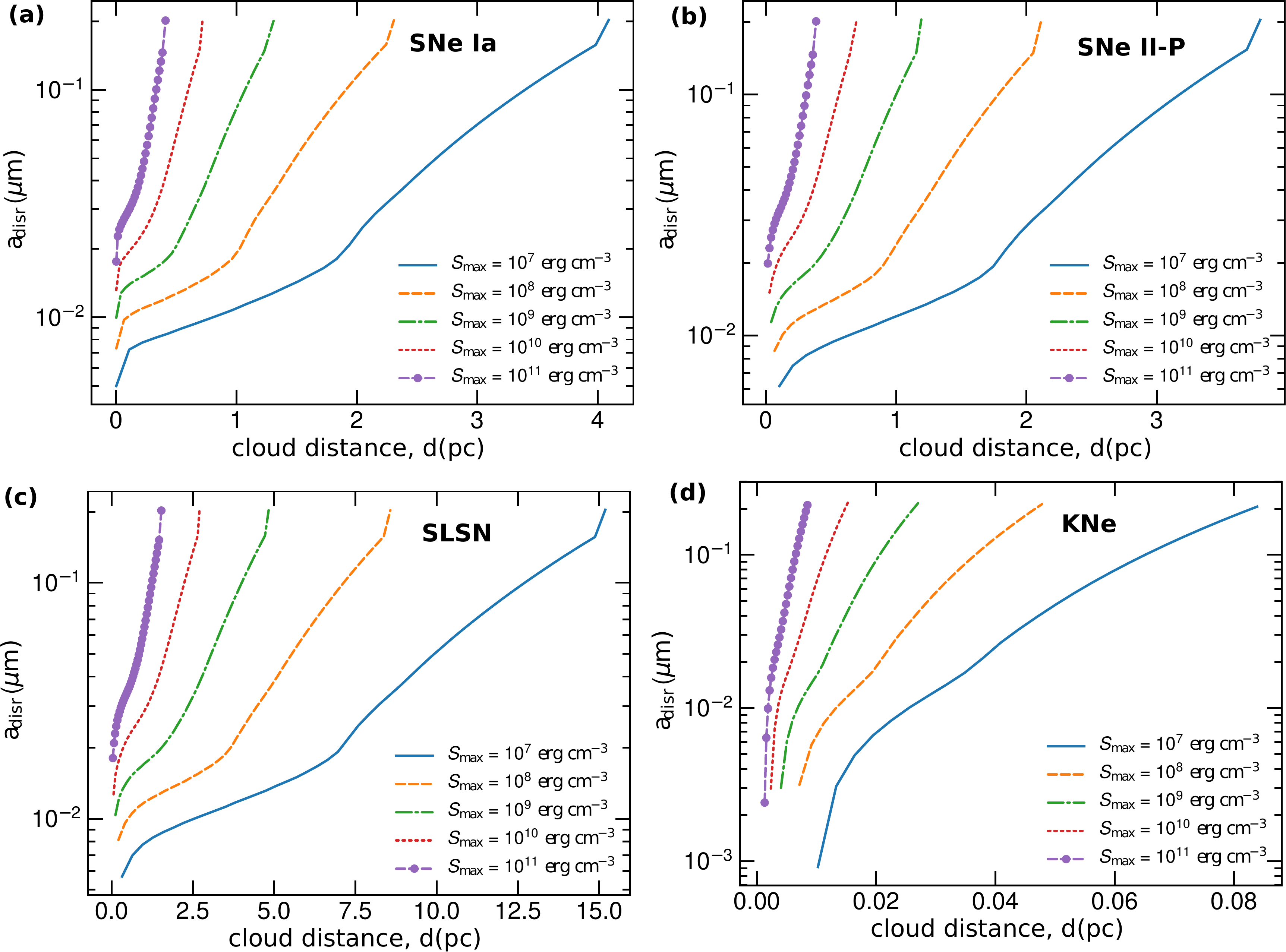}
\caption{{\bf Grain disruption size vs. cloud distance for the time-varying radiation sources assuming different tensile strength $S_{\rm max}$}. Panels ({\bf a})-({\bf d}) show results obtained for SNe Ia, SNe II-P with $L=L_{\rm SNII}$, SLSN with $L=16L_{\rm SNII}$, and KNe. The disruption size increases rapidly with increasing cloud distance, but it decreases with decreasing the grain tensile strength.}
\label{fig:adisr_d_SNe}
\end{figure*}

Figure \ref{fig:adisr_d_SNe} (panels (a)-(d)) shows $a_{\rm disr}$ as a function of the cloud distance for SNe and KNe, assuming a range of the grain tensile strength from $S_{\rm max}= 10^{7}-10^{11}~\rm erg~cm^{-3}$. {For a given $S_{\rm max}$, $a_{\rm disr}$ increases rapidly with the cloud distance due to the decrease of radiation strength $U$ with $d$. Moreover, for a fixed distance, the disruption size increases rapidly with $S_{\rm max}$. The reason is that larger grains experience stronger RATs and rotate faster (see equation (\ref{eq:omega_RAT})) such that the centrifugal stress can exceed the maximum tensile strength $S_{\rm max}$.}
For $S_{\rm max}=10^{9}~\rm erg~cm^{-3}$ (e.g., compact grains), one obtains $a_{\rm disr}\sim 0.1~\mu$m for a dust cloud at a distance of $d\sim 1$ pc. For a lower strength of $S_{\rm max}= 10^{7}~\rm erg~cm^{-3}$ (e.g., composite grains; see Supplementary Table 1), large grains can be disrupted at large distances of $d\sim 4$ pc. The results for SNe II-P are similar to those of SNe Ia, but superluminous supernovae (SLSN) can destroy large grains to farther distances due to their stronger radiation intensity, as expected. The disruption by KNe is efficient within a short distance of $d\lesssim 0.1$ pc due to their lower luminosity.

\begin{figure*}
\centering
\includegraphics[scale=0.45]{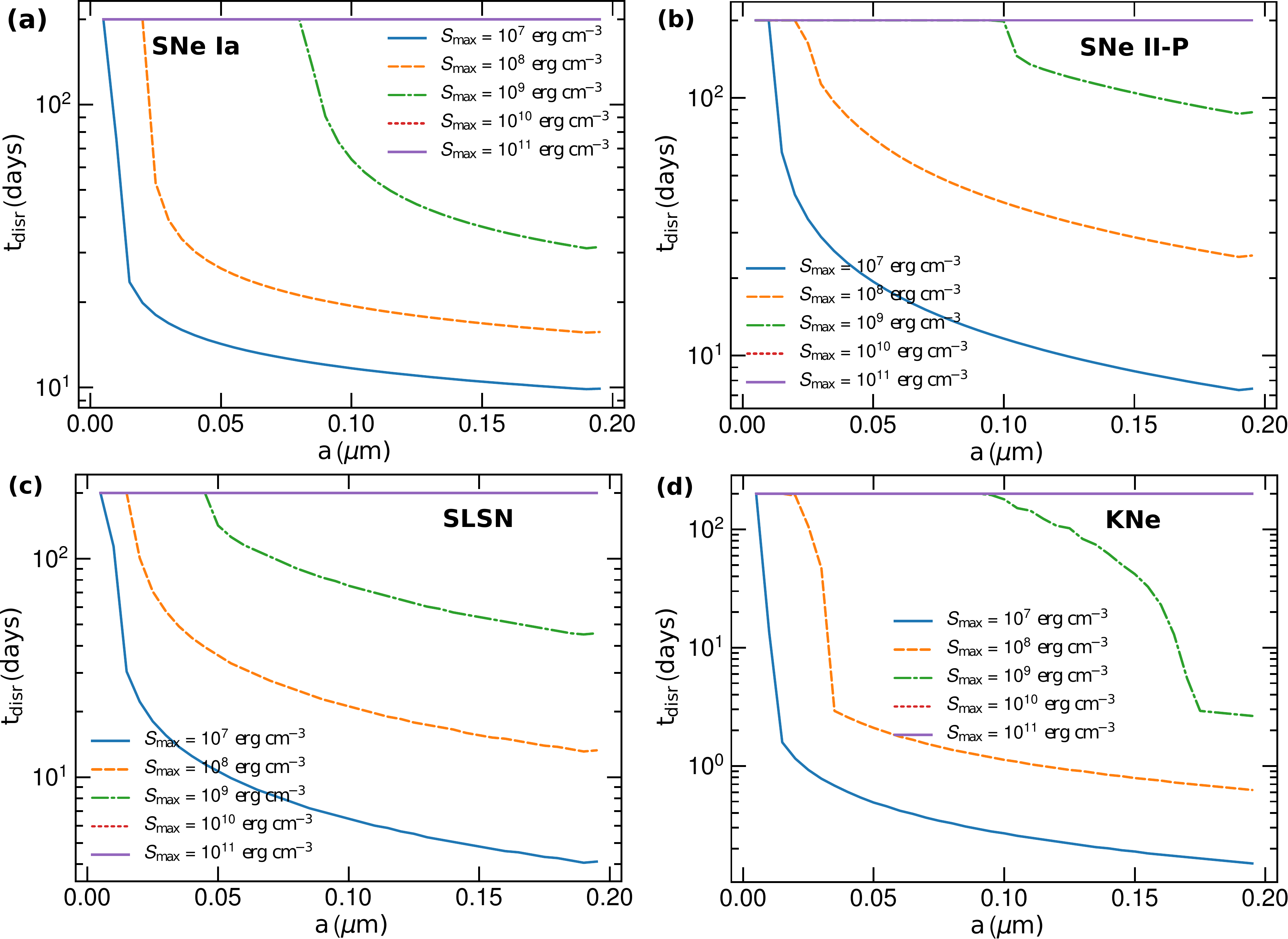}
\caption{{\bf Disruption time vs. grain size for the time-varying radiation sources assuming different tensile strength $S_{\max}$}. Panels ({\bf a})-({\bf d}) show results for SNe Ia, SNe II-P with $L=L_{\rm SNII}$ for dust cloud at 1 pc, SLSN with $L=16L_{\rm SNII}$ for dust cloud at 3 pc, and KNe with dust cloud at 0.02 pc. The disruption time decreases rapidly with increasing grain size and with decreasing tensile strength. Large grains can be rapidly destroyed within 100 days, but the smallest grains are not disrupted and $t_{\rm disr}$ is set to be 200 days.}
\label{fig:tdisr_a_SNe}
\end{figure*}

Figure \ref{fig:tdisr_a_SNe} (panel (a)-(d)) shows the disruption time $t_{\rm disr}$ as a function of the disruption size $a_{\rm disr}$ for SNe, SLSN, and KNe. When the grain size increases to the critical value $a_{\rm disr}$, the disruption time starts to decrease rapidly with increasing the grain size due to stronger RATs (see equation (\ref{eq:tdisr}) in Methods). Large grains ($a\gtrsim 0.1~\mu$m) are destroyed within 50 days from the SNe Ia explosion for $S_{\rm max}=10^{9}~\rm erg~cm^{-3}$, and the disruption time becomes shorter, within 10 days, for $S_{\rm max}=10^{7}~\rm erg~cm^{-3}$ (blue line, panel (a)). The disruption time for SLSN (panel (c)) is shorter than SNe Ia and SN II-P due to its higher luminosity. 

Table \ref{tab:adisr_ISM} lists the grain disruption size and disruption time (estimated for a typical grain size of $a=0.1~\mu$m) for the different cloud distances illuminated by the various radiation fields.

\section*{\large Discussion}\label{sec:dis}
In Table \ref{tab:destr}, we compare the characteristic timescale of RATD with that of other destruction mechanisms, including grain-grain collisions and non-thermal sputtering induced by supernova radiation pressure\cite{2017ApJ...836...13H} (see Supplementary Information section `Comparison of RATD to other destruction mechanisms'). The RATD mechanism is obviously far more efficient than other mechanisms in destroying large grains (i.e., $a\gtrsim 0.1~\mu$m) in strong radiation fields of radiation strength $U\gg 1$.

\subsection{RATD reproduces small grains in H\,{\sc ii} regions around YMSCs.}
\begin{figure*}
\centering
\includegraphics[width=0.9\textwidth]{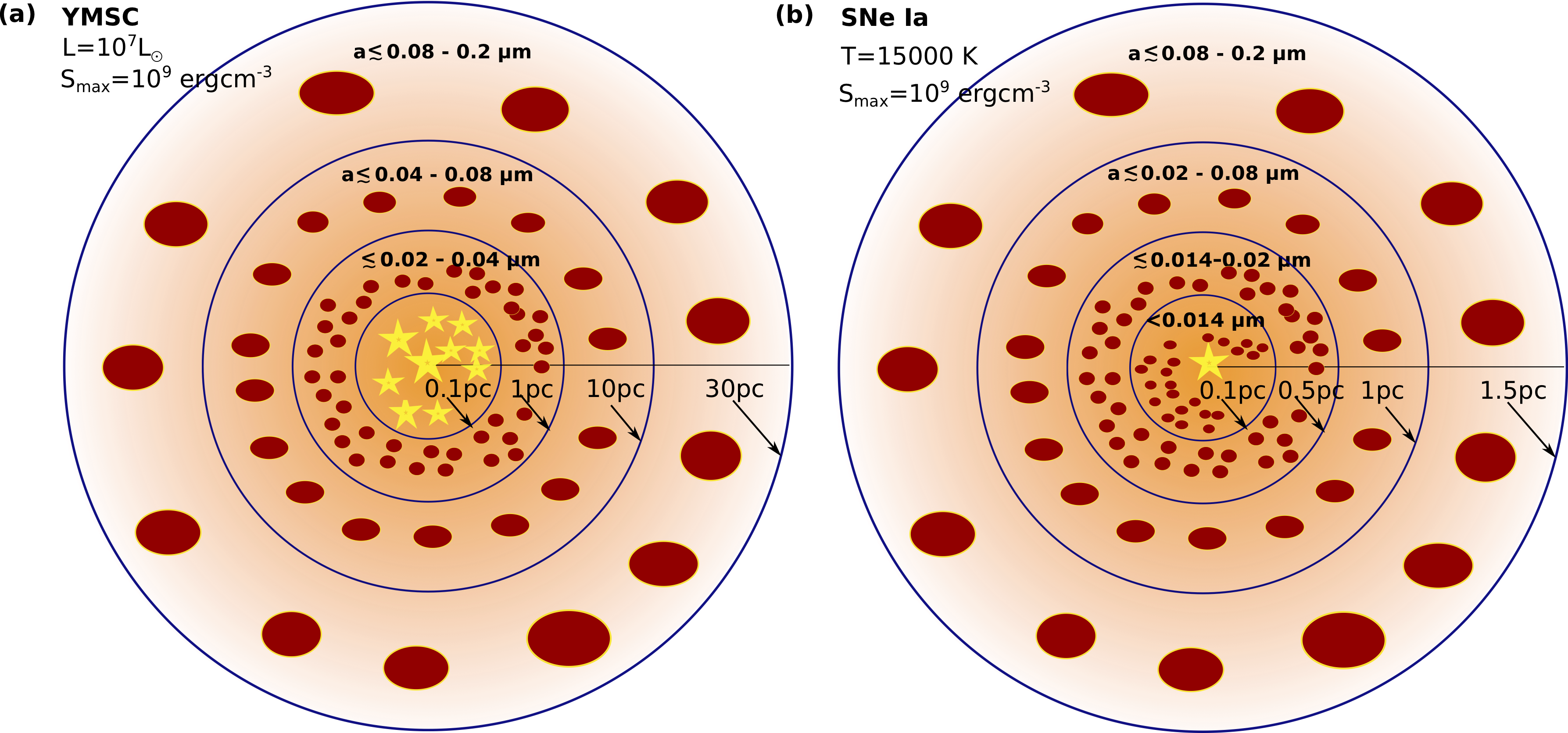}
\caption{{\bf Schematic illustration for the properties of dust grains in a cloud surrounding an intense radiation source modified by RATD, assuming that original dust grains have the same size}. Panel ({\bf a}): illustration for the case of a young massive star cluster (YMSC) with $L=10^{7}L_{\odot}$. Within the distance $d\sim 0.1-1$ pc, only small grains of $a\lesssim 0.02-0.04~\mu$m can survive while larger grains are disrupted by RATD (see the main article for more details). Panel ({\bf b}): illustration for the case of SNe Ia observed at early times (within a few weeks after explosion). Within the distance of $d\sim 0.5-1$ pc, only small grains $a\lesssim 0.02-0.08~\mu$m are present, while larger grains are disrupted. The density of small grains increases toward the center due to disruption of large grains. Here the tensile strength $S_{\rm max}=10^{9}~\rm erg~cm^{-3}$ is assumed.}
\label{fig:dust_HII_SNe}
\end{figure*}

Using the numerical results presented in Fig. \ref{fig:atdisr_dis}, panel (a) and in Figure \ref{fig:dust_HII_SNe} (panel (a), we illustrate the expected range of grain sizes and grain number density present in a H\,{\sc ii} region surrounding a YMSC of luminosity $L=10^{7}L_{\odot}$ as a result of RATD. Large grains of $a> 0.2~\mu$m are destroyed within a large distance of $d\sim 30$ pc from the source, producing smaller grains of $a\lesssim 0.08-0.2~\mu$m within a shell of $d\sim 10-30$ pc. Similarly, RATD destroys large grains of $a> 0.08~\mu$m within a distance of 10 pc, increasing the abundance of small grains of $a\lesssim 0.04-0.08~\mu$m in a shell of $d\sim 1-10$ pc. Within $1$ pc from the source, all large grains are destroyed to further increase the abundance of small grains of $a\lesssim 0.04~\mu$m. A more luminous cluster of $L= 10^{9}L_{\odot}$ can destroy grains of $a\gtrsim 0.02~\mu$m within $\sim$ 1 pc (see Fig. \ref{fig:atdisr_dis} (panel (a)). Therefore, due to the destruction of large grains by RATD, the abundance of small grains is predicted to increase toward the central cluster, as illustrated in Figure \ref{fig:dust_HII_SNe} panel (a).

The reproduction of small grains by RATD can explain the NIR--MIR emission excess observed in H\,{\sc ii} regions around YMSCs\cite{2013A&A...552A.140R,2017ApJ...843...95M}. If large grains are constantly formed in dense regions (e.g., stellar winds and supernova ejecta) and injected into the ISM, then, it takes less than 1 Myr (see Fig. \ref{fig:atdisr_dis}, panel (b)) for RATD to destroy these grains, reproducing small grains. Moreover, RATD can explain the increase in the abundance of small grains relative to large grains toward the cluster center as observed in H\,{\sc ii} regions around massive stars\cite{2014ApJ...784..147S} and YMSCs (NGC 3603)\cite{2007ApJ...665..390L,Relano:2018kx}. 

\subsection{RATD enhances small grains in the regions surrounding supernovae.}
Figure \ref{fig:dust_HII_SNe} (panel (b)) shows a schematic illustration of the size range and density of dust grains in a cloud surrounding a SN Ia due to RATD. The structure is similar to YMSCs, but the extent is smaller due to the shorter shining time of SNe Ia. Large grains of $a\sim 0.08-0.2~\mu$m can be present within a shell of $d\sim 1-1.5$ pc. However, these large grains are destroyed within $d\sim 0.5-1$ pc to increase the abundance of small grains of $a\lesssim 0.02-0.08~\mu$m. Below $d\sim 0.5$ pc, large and small grains are destroyed, such that only very small grains of $a<0.02~\mu$m are present. 

Numerous works suggest that the anomalous values of $R_{\rm V}$ observed toward SNe Ia\cite{2008A&A...487...19N,2010AJ....139..120F} are produced by the predominance of small grains in the host galaxy\cite{2013ApJ...779...38P}.

The inverse modeling for both extinction and polarization curves demonstrates that grains are essentially small, with the peak mass distribution at $a_{\rm peak}\sim 0.07-0.08~\mu$m for SN 1986G, 2006X, 2014J (ref. \cite{2017ApJ...836...13H}). A detailed modeling of extinction curve for SN 2014J (ref. \cite{2015ApJ...807L..26G}) yields $a_{\rm peak}\sim 0.04-0.06~\mu$m. Moreover, the maximum cutoff of the power-law size distribution of $a_{\rm max}\sim 0.094~\mu$m and $0.057~\mu$m are inferred for SNe Ia with $R_{V}=1.5$ and $R_{V}=1$ (ref. \cite{2016P&SS..133...36N}). One can see from Figure \ref{fig:dust_HII_SNe} (panel (b)) that the RATD mechanism can successfully produce such small grains if the dust cloud is located at distance $d\sim 1$ pc, assuming compact grains of $S_{\rm max}= 10^{9}~\rm erg~cm^{-3}$. For composite grains of $S_{\rm max}= 10^{7}~\rm erg~cm^{-3}$, the dust cloud that contains such small grains can be farther from the explosion, at $d\sim 3$ pc.

In the RATD paradigm, large grains ($a\gtrsim 0.1~\mu$m) in a cloud of distance $d<2-3$ pc are destroyed first by RATD, followed by destruction of smaller grains (see Fig. \ref{fig:tdisr_a_SNe}). As a result, the colour excess ($E(B-V)$) of SNe Ia is expected to increase with time due to the increase of small grains if there are dust clouds located within $d<2-3$ pc. This poses a critical challenge for inferring accurate intrinsic light curves of SNe Ia because SNe Ia also experience intrinsic colour variation\cite{2008ApJ...675..626W}. 

The time-varying colour excess is already found toward many highly reddened SNe Ia with anomalous $R_{V}$\cite{2018MNRAS.tmp.1547B}. The authors estimated the distance of dust clouds $d\sim 0.7- 18$ pc for 13 out of the 15 SNe Ia with time-varying $E(B-V)$. Note that their cloud distances are estimated assuming the standard grain size distribution of Milky Way. The inclusion of the smaller grain size cutoff due to RATD is expected to reduce the cloud distance (M. Bulla, private communication). 

\subsection{RATD can modify dust properties around kilonovae.}
In Figures \ref{fig:adisr_d_SNe} and \ref{fig:tdisr_a_SNe}, we showed that large grains within 0.1 pc from the merger location can be destroyed by KNe through RATD. Therefore, the surrounding environments should be dominated by small grains. The time-varying colour excess due to RATD would be a critical problem for deriving the intrinsic colour of KNe because the latter also varies with time\cite{2012ApJ...746...48M}.

While large grains can be destroyed by KNe, small grains receiving weaker torques can be spun-up to suprathermal rotation and be aligned with the magnetic field by radiative torques\cite{2017ApJ...836...13H}. Polarized dust emission from aligned grains can be used to trace magnetic fields in the environment where the merger occurs.

Recently, the properties of the ISM around gravitational wave (GW) source (GW170817) were studied, where a low value of $E(B-V)=0.08$ was reported, assuming the typical $R_{\rm V}=3.1$ (ref. \cite{Levan:2017jz}). If there is some dust cloud in the vicinity of the KN, then, one expects a lower value of $R_{\rm V}$ due to the excess of small grains produced by RATD, resulting in a larger colour excess. Finally, polarimetric observations of GW170817 reveal a low degree of polarization\cite{Covino:2017dq}, suggesting that most polarization is produced by aligned grains in the interstellar medium rather than from scattering in the ejecta. Therefore, polarimetric observations of KNe are useful for probing dust properties around GW sources.

\subsection{Implications of RATD for other astrophysical conditions.}

In starburst galaxies with star formation activities and supernova explosions, observations usually show peculiar extinction curves with a steep far-UV rise\cite{1997ApJ...487..625G}. Such a feature is usually explained by grain shattering due to supernova shocks. However, this mechanism requires a long timescale to be efficient and cannot explain the absence of $2175$\AA~ extinction bump presumably produced by very small carbonaceous grains\cite{2003ARA&A..41..241D}.

Here we suggest that the steep far-UV rise in starburst galaxies can be produced by the excess of small grains arising from RATD in intense radiation of massive stars and supernovae. Moreover, it is well-known that tiny grains are efficiently destroyed by extreme UV and X-rays of intense radiation fields. Therefore, the RATD mechanism can successfully explain both the observed steep far-UV rise and the lack of $2175$~\AA~ extinction bump.

The RATD mechanism can also have implications for the study of the escape fraction of Lyman-$\alpha$ photons, $f_{\rm esc}$, from star-forming galaxies. As shown in Figure \ref{fig:dust_HII_SNe} (panel (a)), dust grains of $a\gtrsim 0.08~\mu$m can be disrupted to a distance of 10 pc in the intense radiation field of YMSCs. The resulting enhancement of small grains will increase the far-UV dust extinction, which can reduce $f_{\rm esc}$, as previously observed\cite{2017ApJ...844..171Y} (see Methods section `Effect of RATD on Ly$\alpha$ photon escape from star-forming galaxies').

Last but not least, the RATD mechanism naturally works in other intense radiation fields such as those around red giant stars, AGNs, quasars, and gamma ray burst (GRB) afterglows, resulting in the excess of small grains in their environs. Therefore, the RATD mechanism can explain the origin of a steep far-UV rise in the SMC-like extinction curves toward GRBs host galaxies\cite{2012A&A...537A..15S}, quasars\cite{2004AJ....128.1112H}, and high-z star-forming galaxies\cite{2018ApJ...853...56R}.

\section*{\large Conclusions}\label{sec:summ}
In this paper, we have introduced a new mechanism of dust destruction that is based on the fact that extremely fast rotating grains spun-up by radiative torques would be disrupted when the centrifugal stress exceeds the maximum tensile strength of grains. This Radiative Torque Disruption (RATD) mechanism is shown to be more efficient than previously-known mechanisms in destroying large grains in strong radiation fields.

Using the RATD mechanism, we quantified the critical size of rotational disruption for grains and its required disruption time for YMSCs, supernovae, and kilonovae. We find that the RATD mechanism can rapidly replenish small grains as required to reproduce the NIR--MIR emission excess as well as the increase in abundance of small grains toward the center observed in YMSCs.

We show that RATD can destroy large grains of $a\gtrsim 0.1~\mu$m within distance of $d\sim 1$ pc (compact grains) and $\sim 3$ pc (composite grains) from SNe Ia and SNe II-P, which can successfully explain the excessive abundance of small grains toward SNe as demonstrated by the anomalous dust extinction and polarization.

Using the RATD mechanism, we study the feedback of kilonovae on surrounding dust and predict that the environments in the vicinity of the mergers would be dominated by small grains. 

Finally, we suggest that the steep far-UV rise in the extinction curves of star-forming galaxies, quasars, and high-z galaxies can be explained by the enhanced abundance of small grains due to grain disruption. The low escape fraction of Ly$\alpha$ photons from star-forming galaxies can also originate from the enhanced dust extinction due to an excess of small grains as a result of RATD.
 
\newpage

\section*{\large Methods}\label{sec:method}
\subsection{Radiation fields and radiative torques of irregular grains.}\label{sec:spinup}
Let $u_{\lambda}$ be the spectral energy density of a radiation field. The energy density of the radiation field is then $u_{\rm rad}=\int u_{\lambda}d\lambda$. The radiation strength is characterized by $U=u_{\rm rad}/u_{\rm ISRF}$ with $u_{\rm ISRF}=8.64\times 10^{-13}~\rm erg~cm^{-3}$ being the energy density of the average interstellar radiation field (ISRF) in the solar neighborhood\cite{1983A&A...128..212M}. 

For a point source of radiation field with bolometric luminosity $L$, the radiation energy density at distance $d_{\rm pc}$ in units of pc is given by
\begin{eqnarray}
u_{\rm rad}=\int u_{\lambda}d\lambda =\int  \frac{L_{\lambda}e^{-\tau_{\lambda}}}{4\pi c d^{2}}d\lambda
\simeq1.06\times 10^{-6}\left(\frac{L_{9}e^{-\tau}}{d_{\rm pc}^{2}}\right) \rm erg~ cm^{-3},\label{eq:urad}
\end{eqnarray}
where  $L_{9}=L/(10^{9}L_{\odot})$, $\tau_{\lambda}$ is the optical depth induced by intervening dust, and $\tau$ is defined as $e^{-\tau}=\int L_{\lambda}e^{-\tau_{\lambda}}d\lambda/L$. The radiation strength is then 
\begin{eqnarray}
U_{6}=\frac{U}{10^{6}}\simeq 1.2\left(\frac{L_{9}e^{-\tau}}{d_{\rm pc}^{2}}\right).
\end{eqnarray}
Thus, at distance $d=1$ pc, a massive star of $L\sim 10^{5}L_{\odot}$ can produce radiation strength of $U\sim 10^{2}$ and a supernova of $L\sim 10^{9}L_{\odot}$ gives $U\sim 10^{6}$.

Dust grains of irregular shape subject to an anisotropic radiation field experience radiative torques (RATs) due to differential scattering and absorption of incident photons\cite{1976Ap&SS..43..291D}. Numerical calculations of RATs for several irregular shapes\cite{1996ApJ...470..551D} found that grains can be spun-up to suprathermal rotation. The spin-up by RATs was subsequently tested in lab experiments\cite{2004ApJ...614..781A}. An analytical model of RATs and RAT alignment was developed\cite{2007MNRAS.378..910L,Hoang:2008gb}, which is supported by numerous observations\cite{Andersson:2015bq,LAH15}.

Let $a$ be the effective size of the irregular grain which is defined as the radius of an equivalent sphere of the same volume as the irregular grain. The radiative torque of the irregular grain produced by the illumination of the anisotropic radiation is written as
\begin{eqnarray}
{\Gamma}_{\lambda}=\pi a^{2}
\gamma u_{\lambda} \left(\frac{\lambda}{2\pi}\right){Q}_{\Gamma},\label{eq:GammaRAT}
\end{eqnarray}
where $\gamma$ is the anisotropy degree of the radiation field, and ${Q}_{\Gamma}$ is the RAT efficiency\cite{1996ApJ...470..551D,2007MNRAS.378..910L}.

The value of $Q_{\Gamma}$ depends on the radiation wavelength, grain shape, and grain size, but it seems to weakly depend on the grain composition and internal structure\cite{2007MNRAS.378..910L,Herranen:2018wd}. The RAT efficiency can be approximately described by a power law\cite{2007MNRAS.378..910L}:
\begin{eqnarray}
Q_{\Gamma}\sim 0.4\left(\frac{{\lambda}}{1.8a}\right)^{\eta},\label{eq:QAMO}
\end{eqnarray}
where $\eta=0$ for $\lambda \lesssim 1.8a$,  and $\eta=-3$ for $\lambda > 1.8a$. 

Let $\bar{\lambda}=\int \lambda u_{\lambda}d\lambda/u_{\rm rad}$ be the mean wavelength of the radiation spectrum. The RAT efficiency averaged over the radiation spectrum is defined as
\begin{eqnarray}
\overline{Q}_{\Gamma} = \frac{\int \lambda Q_{\Gamma}u_{\lambda} d\lambda}{\int \lambda u_{\lambda} d\lambda}.
\end{eqnarray}

For dust grains of size $a\lesssim \bar{\lambda}/1.8$, the averaged RAT efficiency, $\overline{Q}_{\Gamma}$, can be approximated to\cite{2014MNRAS.438..680H}
\begin{eqnarray}
\overline{Q}_{\Gamma}\simeq 2\left(\frac{\bar{\lambda}}{a}\right)^{-2.7}\simeq 2.6\times 10^{-2}\bar{\lambda}_
{0.5}^{-2.7}a_{-5}^{2.7},
\end{eqnarray}
where $a_{-5}=a/(10^{-5}\rm cm)$, $\bar{\lambda}_{0.5}=\bar{\lambda}/(0.5~\mu\rm m)$, and $\overline{Q}_{\Gamma}\sim 0.4$ for $a> \bar{\lambda}/1.8$.

As a result, the averaged radiative torque can be written as
\begin{eqnarray}
\Gamma_{\rm RAT}=\pi a^{2}
\gamma u_{\rm rad} \left(\frac{\bar{\lambda}}{2\pi}\right)\overline{Q}_{\Gamma}\simeq 5.6\times 10^{-23}\gamma a_{-5}^{4.7}\bar{\lambda}_
{0.5}^{-1.7}U_{6}~\rm erg
\end{eqnarray}
for $a\lesssim \bar{\lambda}/1.8$, and 
\begin{eqnarray}
\Gamma_{\rm RAT}&\simeq & 8.6\times 10^{-22}\gamma a_{-5}^{2}\bar{\lambda}_{0.5}U_{6}~\rm erg
\end{eqnarray}
for $a> \bar{\lambda}/1.8$.

\subsection{Maximum grain angular velocity spun-up by radiative torques.}

RATs can spin-up dust grains to fast rotation. However, faster grain rotation induces faster rotational damping due to both gas collisions and emission of infrared photons. The equation of motion for irregular grains subject to RATs and rotational damping is described by
\begin{eqnarray}
\frac{Id\omega}{dt} = \Gamma_{\rm RAT}-\frac{I\omega}{\tau_{\rm damp}},\label{eq:domega}
\end{eqnarray}
where $I=8\pi \rho a^{5}/15$ is the grain inertia moment, and
\begin{eqnarray}
\tau_{\rm damp}^{-1}=\tau_{\rm gas}^{-1}(1+ F_{\rm IR}),\label{eq:taudamp}
\end{eqnarray}
where $\tau_{\rm gas}$ is the characteristic rotational damping time due to gas collisions (see Supplementary section `Grain rotational damping') :
\begin{eqnarray}
\tau_{\rm gas}&=&\frac{3}{4\sqrt{\pi}}\frac{I}{1.2n_{\rm H}m_{\rm H}
v_{\rm th}a^{4}}\nonumber\\
&\simeq& 8.74\times 10^{4}a_{-5}\hat{\rho}\left(\frac{30~\rm cm^{-3}}{n_{\rm H}}\right)\left(\frac{100~\rm K}{T_{\rm gas}}\right)^{1/2}~{\rm yr},\label{eq:taudamp_gas}
\end{eqnarray}
where $v_{\rm th}=\left(2k_{\rm B}T_{\rm gas}/m_{\rm H}\right)^{1/2}$ is the thermal velocity of hydrogen atoms of mass $m_{\rm H}$\cite{1996ApJ...470..551D}. 

In equation (\ref{eq:taudamp}), $F_{\rm IR}$ is a dimensionless parameter that describes the relative importance of rotational damping by infrared (IR) emission with respect to the damping by gas collisions, as given by (see ref. \cite{1998ApJ...508..157D}):
\begin{eqnarray}
F_{\rm IR}\simeq \left(\frac{0.4U^{2/3}}{a_{-5}}\right)
\left(\frac{30~ \rm cm^{-3}}{n_{\rm H}}\right)\left(\frac{100~ \rm K}{T_{\rm gas}}\right)^{1/2}.\label{eq:FIR}
\end{eqnarray} 

For the radiation source of constant luminosity (e.g., massive stars or YMSCs), $\Gamma_{\rm RAT}$ is constant. From equation (\ref{eq:domega}), one can solve for the grain angular velocity as a function of time:
\begin{eqnarray}
\omega(t)=\omega_{\rm RAT}\left[1-\exp\left(-\frac{t}{\tau_{\rm damp}}\right)\right],\label{eq:omega_time}
\end{eqnarray}
where
\begin{eqnarray}
\omega_{\rm RAT}=\frac{\Gamma_{\rm RAT}\tau_{\rm damp}}{I}\label{eq:omega_RAT0}
\end{eqnarray}
is the terminal angular velocity at $t\gg \tau_{\rm damp}$, which is considered the maximum rotational rate spun-up by RATs.

Equation (\ref{eq:omega_RAT0}) allows us to calculate the maximum rotation rate enabled by RATs for arbitrary environments with given $n_{\rm H}, T_{\rm gas}$, and $U$. For the case of strong radiation fields with $U\gg 1$, IR damping is dominant over gas damping (i.e.,  $F_{\rm IR}\gg 1$), such that $\tau_{\rm damp}\simeq \tau_{\rm gas}F_{\rm IR}$. As a result, by plugging $\tau_{\rm damp}$ and $\Gamma_{\rm RAT}$ into equation (\ref{eq:omega_RAT0}), we can derive analytical formulae for $\omega_{\rm RAT}$ as given by equations (\ref{eq:omega_RAT}) and (\ref{eq:omega_RAT1}).

For the time-varying radiation sources such as supernovae and kilonovae, the grain angular velocity can be obtained by numerically solving the equation of motion:
\begin{eqnarray}
Id\omega = \left(\Gamma_{\rm RAT}(t)-\frac{I\omega}{\tau_{\rm damp}}\right)dt,\label{eq:dodt}
\end{eqnarray}
where the initial conditions $\omega=0$ at $t=0$ can be chosen.

We note that equations (\ref{eq:omega_time})-(\ref{eq:dodt}) can be used for specific shapes of irregular grains\cite{1997ApJ...480..633D,2007MNRAS.378..910L,Herranen:2018wd} by replacing $I$ with the principal inertia moment along the axis of maximum inertia. The reason is that, for suprathermally rotating grains spun-up by intense radiation, internal relaxation can rapidly dissipate the grain rotational energy to the minimum level, resulting in the perfect alignment of the axis of maximum inertia moment with the grain angular momentum\cite{1979ApJ...231..404P}.

\subsection{Centrifugal stress within a spinning grain.}\label{sec:stress}
For an equivalent spherical grain spinning around a $z-$ axis with angular velocity $\omega$, the centrifugal stress on a circular slab of thickness $dx$ located at distance $x_{0}$ from the mass center is equal to
\begin{eqnarray}
dS=\frac{\omega^{2}x dm}{\pi (a^{2}-x_{0}^{2})}= \frac{\rho \omega^{2}(a^{2}-x^{2})xdx}{a^{2}-x_{0}^{2}},
\end{eqnarray}
where the mass of the slab $dm=\rho dAdx$ with $dA=\pi(a^{2}-x^{2})$ the area of the circular slab.

The average centrifugal stress is then given by
\begin{eqnarray}
S&=&\int_{x_{0}}^{a} dS=\frac{\rho \omega^{2}a^{2}}{2}\int_{x_{0}/a}^{1}\frac{(1-u)du}{1-u_{0}}\nonumber\\
&=& \frac{\rho \omega^{2}a^{2}}{4}\left[\frac{(1-u_{0})^{2}}{1-u_{0}}\right]=\frac{\rho \omega^{2}a^{2}}{4}\left[1-\left(\frac{x_{0}}{a}\right)^{2}\right],~~~\label{eq:Smax}
\end{eqnarray}
where $u=x^{2}/a^{2}$. The centrifugal stress is maximum of $S=\rho \omega^{2}a^2/4$ for $x_{0}=0$.

\subsection{Radiation fields of massive stars, supernovae, and kilonovae}
Massive OB stars can produce strong radiation fields of $L\sim 10^{3}-10^{6}L_{\odot}$. A young massive star cluster (hereafter YMSC) contains more than thousands of OB stars and more than $10^{4}M_{\odot}$, and its luminosity spans between $10^{6}-10^{9}L_{\odot}$ (ref. \cite{PortegiesZwart:2010kc}). 

Massive stars can be considered as black-body radiation sources, such that its radiation spectrum can be described by the Planck function, assuming an effective temperature $T_{\star}= 20000~\rm K$. The mean wavelength of the radiation spectrum is then 
\begin{eqnarray}
\bar{\lambda}=\frac{\int \lambda u_{\lambda}(T_{\star}) d\lambda}{\int u_{\lambda}(T_{\star})d\lambda} \approx 0.28~\mu \rm m.
\end{eqnarray}
Thus, for massive stars and YMSCs, we can assume a constant radiation field over the lifetime of massive stars with the anisotropy degree of radiation $\gamma=1$. 

Supernova explosions can release radiation energy of $L\sim 10^{8}-10^{10}L_{\odot}$ over a timescale of $\sim 100-150$ days. Radiation energy of SNe Ia is produced mostly by conversion of the kinetic energy of ejecta interacting with surrounding environments (i.e., shocked regions). Most of radiative energy is concentrated in UV-optical wavelengths, especially in early epochs after the explosion\cite{2009AJ....137.4517B}.

An analytical formula for fitting optical luminosity of the SNe Ia light curve is introduced\cite{2017ApJ...848...66Z}. The formula appears to fit well the B and g bands during the first 50 days\cite{2017ApJ...848...66Z}. For RATs, we are interested only in UV-optical wavelengths, thus, we can use their analytical fit for our numerical calculations of grain rotation rate, which consists of two power laws:
\begin{eqnarray}
L_{\rm SNIa}(t) = L_{0}\left(\frac{t-t_{0}}{t_{b}}\right)^{\alpha_{r}}\left[1+\left(\frac{t-t_{0}}{t_{b}}\right)^{s\alpha_{d}}\right]^{-2/s},~~~\label{eq:LSN}
\end{eqnarray}
where $L_{0}$ is the scaling parameter, the first power law describes the rising period in the luminosity\cite{1999AJ....118.2675R}, and the second power law (square brackets) describes the falling stage after the peak luminosity. The parameters $\alpha_{r}$ and $\alpha_{d}$ are the power law indices before and after the peak, $t_{0},t_{b}$ is the first- light time and the break time, and $s$ is the smooth transition parameter. We adopt $\alpha_{r}\sim 2$, $\alpha_{d}\sim 2.5$, $s\sim 1$, and  $t_{b}\sim 23$ days, which are the best-fit parameters\cite{2017ApJ...848...66Z}, and $L_{0}=2\times 10^{10}L_{\odot}$. For convenience, we set the first-light time $t_{0}=0$, so that $t$ corresponds to the time interval that dust grains in a nearby cloud are illuminated by supernova radiation.

We now consider the case of Type II-P supernovae, a special class of Type II SNe with an extended plateau in the light curve. We use the arbitrary power-law fits for the light curve of the supernova 2015ba (ref. \cite{2018MNRAS.479.2421D}), which consists of three components:
\begin{eqnarray}
    L_{\rm SNII}(t)= \left\{
\begin{array}{ll}
    10^{-0.002t+9.17}L_{\odot}, {~\rm for~ } t\leq 60\ {\rm days}\\
    1.1\times 10^{9}L_{\odot}, {~\rm for~} 60<t\leq 120\ {\rm days}\\
    10^{-0.0067t+9.84}L_{\odot}, {~\rm for~} t>120 {~\rm days.}
\end{array}
	\right. ~~~\label{eq:LSNII}
\end{eqnarray}

Recently, a new class of supernovae, namely superluminous supernovae (SLSN), was discovered\cite{2012Sci...337..927G}. The luminosity can be two orders of magnitude higher than the standard SNe Ia luminosity\cite{2016Sci...351..257D}. For the sake of completeness, we calculate the disruption size and disruption time for a case of SLSN that has the peak luminosity 16 times larger than $L_{\rm SNII}$ given by equation (\ref{eq:LSNII}). 

Kilonova is electromagnetic radiation mostly in UV, optical and near infrared, which is produced by the decay of radioactive nuclei in the relativistic ejecta after the merger of black hole (BH)-neutron star (NS) and NS-NS stars\cite{2010MNRAS.406.2650M,2017PASJ...69..102T}.

The bolometric luminosity of kilonovae can reach the peak of $L_{\rm peak}\sim 3\times 10^{8}L_{\odot}$ after 0.5-5 days from the merger moment. After the peak, the luminosity decreases with time as
\begin{eqnarray}
L_{\rm KN}(t)=L_{\rm peak}t^{-\alpha},\label{eq:LKN}
\end{eqnarray}
where the time $t$ is given in units of day, $\alpha\sim 0.2$ for $t\le 3$ days (shallow slope), and $\alpha=1.3$ for $t>3$ days (steep slope)\cite{2010MNRAS.406.2650M}.

The radiation spectrum of SNe Ia is also assumed to be black body of effective temperature $T_{\star}$. This is especially valid during the rising phase of the SNe Ia light curve\cite{1999AJ....118.2675R}. Thus, for $T_{\star}=1.5\times 10^{4}$ K, one obtains $\bar{\lambda}\approx 0.35~\mu$m for SNe Ia. For SNe II-P and SLSN, which are assumed to have similar spectrum with $T_{\star}= 2\times 10^{4}$ K but the luminosity of SLSN is 16 times larger, one obtains $\bar{\lambda}\approx 0.28~\mu$m. The radiation spectrum of KNe can also be approximated by a temperature $T_{\star}= 10^{4}$ K\cite{2012ApJ...746...48M}, which yields $\bar{\lambda}\approx 0.53~\mu$m.

\subsection{Critical radiation strength for RATD.}
For strong radiation fields for which IR damping dominates over the gas damping (i.e., $F_{\rm IR}\gg 1$), from equations (\ref{eq:omega_RAT}) and (\ref{eq:omega_cri}), one can derive the minimum radiation strength required to disrupt irregular grains of $a\lesssim \bar{\lambda}/1.8$ as follows:
\begin{eqnarray}
U\ge U_{\rm disr} \simeq 
	118\gamma^{-3}a_{-5}^{-8.1}\bar{\lambda}_{0.5}^{5.1}S_{\rm max,9}^{3/2}.\label{eq:Udisr}
\end{eqnarray}

The above equation reveals that the critical radiation strength depends on the anisotropy of the radiation field $\gamma$, the mean wavelength, the grain size, and especially the grain mechanical property or tensile strength. Compact grains with high $S_{\rm max}$ are more difficult to disrup than fluffy/composite grains. Note that radiative torques from solar radiation may be able to disrupt fluffy, submicron-sized interplanetary grains\cite{2016ApJ...818..133S}. 

For a general case, one can find $U_{\rm disr}$ by comparing $\omega_{\rm RAT}$ obtained from numerical calculations using equation (\ref{eq:omega_RAT0}) with $\omega_{\rm disr}$ (equation (\ref{eq:omega_cri})). Supplementary Figure 1 (panel (a)) shows the critical radiation strength $U_{\rm disr}$ as a function of the grain size for the diffuse ISRF ($\gamma=0.1, \bar{\lambda}=1.2~\mu$m) and unidirectional radiation sources ($\gamma=1, \bar{\lambda}=0.2-0.8~\mu$m). The radiation strength $U_{\rm disr}$ decreases rapidly with increasing $a$ as $a^{-8.1}$, then it starts to rise slowly from $a\sim \bar{\lambda}/1.8$ because the RAT efficiency increases rapidly with $a$ and becomes constant for $a>\bar{\lambda}/1.8$ (see equation (\ref{eq:QAMO})). For the diffuse interstellar medium with $\gamma=0.1$ and $\bar{\lambda}=1.2~\mu$m, it requires $U_{\rm disr}\sim 900$ to disrupt grains $a\sim 0.3~\mu$m. However, for the case of unidirectional radiation from a distant star with $\gamma=1$, one only requires $U_{\rm disr}\sim 0.9$ to disrupt the similar size grain. 

In Supplementary Figure 1 (panel (b)), we show the results computed for the photodissociation region (PDR) conditions with $n_{\rm H}=10^{5}~\rm cm^{-3}$ and $T_{\rm gas}=10^{3}~\rm K$. Due to the enhanced gas damping, the required radiation strength is much larger.

\subsection{Grain disruption size and disruption time.}

For strong radiation fields of $U\gg 1$ such that the gas damping can be disregarded due to the dominance of IR damping, using equations (\ref{eq:omega_RAT}) and (\ref{eq:omega_cri}), one can obtain the critical size $a_{\rm disr}$ above which all grains would be disrupted (i.e., disruption size), for $a_{\rm disr}\lesssim \bar{\lambda}/1.8$, as follows:
\begin{eqnarray}
\left(\frac{a_{\rm disr}}{0.1~\mu\rm m}\right)^{2.7}&\simeq&0.046\gamma^{-1}\bar{\lambda}_
{0.5}^{1.7}\left(\frac{U_{6}}{1.2}\right)^{-1/3}S_{\rm max,9}^{1/2}.~~~\label{eq:a_cri}
\end{eqnarray}

For a cloud at distance $d=1~\rm pc$ and $\bar{\lambda}=0.5~\mu$m, equation (\ref{eq:a_cri}) gives $a_{\rm disr}\sim 0.03~\mu$m for $U= 10^{6}$ and $S_{\rm max}=10^{9}~\rm erg~cm^{-3}$. At $d=10~\rm pc$, one obtains $a_{\rm disr}\sim 0.08~\mu$m for the similar parameters, and $a_{\rm disr}\sim 0.13~\mu$m for $d=50$ pc. We note in dense environments or not strong radiation fields where gas damping is dominant over IR damping, there exists an upper cutoff of the grain disruption sizes by RATD because $\omega_{\rm RAT}$ decreases faster than $\omega_{\rm disr}$ for $a>\bar{\lambda}/1.8$ \cite{Hoang:2019td}.

The disruption time of grains of size $a_{\rm disr}$ can be defined as the time required to spin-up grains to $\omega_{\rm disr}$:
\begin{eqnarray}
t_{\rm disr}=\frac{I\omega_{\rm disr}}{dJ/dt}=\frac{I\omega_{\rm disr}}{\Gamma_{\rm RAT}}
\simeq 368\bar{\lambda}_
{0.5}^{1.7}\left(\frac{a_{\rm disr}}{0.1~\mu\rm m}\right)^{-0.7}U_{6}^{-1}S_{\rm max,9}^{1/2}{~\rm days}.\label{eq:tdisr}
\end{eqnarray}

For a dust cloud at $d=1\rm~ pc$ from a supernova with $U\sim 10^{6}$, the disruption time $t_{\rm disr}\sim 161$ days for $a_{\rm disr}\sim ~0.25~\mu$m, assuming $\bar{\lambda}=0.5~\mu$m. The disruption time decreases with the grain size, such that large grains are disrupted faster than smaller ones.

For SNe and KNe with the time-dependent luminosity, we solve the equation of motion (equation (\ref{eq:domega})) to obtain $\omega(t)$. The disruption size and disruption time is determined by comparing $\omega(t)$ with the critical disruption limit $\omega_{\rm disr}$ (see Supplementary Figure 2).

\subsection{Comparison of RATD to other destruction mechanisms in strong radiation fields}\label{sec:comp}
To facilitate the comparison of RATD to other destruction mechanisms, let us first describe the basic properties of the RATD mechanism. 

The efficiency of the RATD mechanism in destroying dust grains in general depends on the radiation strength ($U$), the anisotropy degree of the radiation field ($\gamma$), local gas properties ($n_{\rm H},T_{\rm gas}$), and maximum tensile strength of dust grains. In the absence of gas rotational damping, the radiation strength required for grain disruption is not too high, such as $U_{\rm disr}< 118\gamma^{-3}S_{\rm max,9}^{3/2}$ for $a\gtrsim 0.1~\mu$m. In the presence of gas damping, $U_{\rm disr}$ increases with increasing the gas rotational damping rate. For a very dense and hot environment like PDRs, the disruption occurs when $10^{4}<U_{\rm disr}<10^{5}$, assuming $\gamma=1$ (see Supplementary Figure 1). The typical radiation strength of PDRs is $U\sim 3\times 10^{4}$, which is sufficient to disrupt large grains. We note that the radiation strength can vary from $U\sim 0.1-10^{7}$ in galaxies\cite{2007ApJ...663..866D}. Therefore, the effect of RATD by strong radiation fields should be accounted for in dust modeling and galaxy evolution research. Moreover, in dense regions with weak radiation fields, such as dense starless cores or the interior of protoplanetary disks, RATD is inefficient. On the other hand, in dense regions but with intense radiation fields such as around young stellar objects, the RATD mechanism can be efficient\cite{Hoang:2019td}.

The efficiency of RATD closely depends on the maximum tensile strength of dust grains, which is determined by internal structures of dust grains. Compact grains tend to have high $S_{\rm max}$ up to $10^{11}~\rm erg~cm^{-3}$, while porous/composite grains have much lower tensile strength (see Supplementary Table 1). If large grains are an aggregate of nanoparticles\cite{1989ApJ...341..808M}, we can expect $S_{\rm max}$ as low as $10^{7}(1-P)~\rm erg~cm^{-3}$ (see Supplementary equation (9)). Therefore, RATD is efficient in disrupting aggregates than compact grains. Furthermore, the efficiency of RATD mechanism is expected to be weakly dependent on grain compositions (i.e., silicate and carbonaceous materials) because RATs are found not vary significantly with grain composition\cite{Herranen:2018wd}.

Now we compare the RATD mechanism with other destruction mechanisms that can work in strong radiation fields, including thermal sublimation, thermal sputtering, and grain shattering due to grain-grain collisions induced by radiation pressure\cite{Hoang:2017kg}. The thermal sublimation is previously known to be the dominant mechanism of dust destruction by intense radiation fields\cite{2000ApJ...537..796W,2015ApJ...806..255H}.

For a point source of radiation, the sublimation distance of dust grains, $r_{\rm sub}$, from the central source is given by
\begin{eqnarray}
r_{\rm sub}\simeq 0.015\left(\frac{L_{\rm UV}}{10^{9}L_{\odot}}\right)^{1/2}\left(\frac{T_{\rm sub}}{180
0\rm K}\right)^{-5.6/2} {\rm pc},\label{eq:rsub}
\end{eqnarray}
where $L_{\rm UV}$ is the luminosity in the optical and UV, which is roughly one half of the bolometric luminosity, and $T_{\rm sub}$ is the dust sublimation temperature between 1500-1800 K for silicate and graphite material\cite{1995ApJ...451..510S,1989ApJ...345..230G}. 

Equation (\ref{eq:rsub}) reveals that thermal sublimation is efficient for grains at much smaller distances compared to RATD (see, e.g., Figure \ref{fig:adisr_d_SNe}). Moreover, at the same distance, the radiation strength required for thermal sublimation of $U_{\rm sub}>8\times 10^{7}(T_{\rm sub}/1800\rm K)^{5.6}$ (see equation (\ref{eq:rsub})), which is much larger than $U_{\rm disr}$. The dominance of rotational disruption over thermal sublimation can be understood by means of energy consideration. Indeed, in order to heat the dust grain to the sublimation temperature, $T_{\rm sub}$, the radiation energy must be $E_{\rm rad}\sim T_{\rm sub}^{4}$. On the other hand, in order to spin-up dust grains to the critical rotation rate $\omega_{\rm disr}$, the radiation energy required is $E_{\rm rad}\sim \omega_{\rm disr}$. Due to the fourth order dependence, the sublimation energy is much higher than the energy required for grain disruption. 

Thermal sputtering is usually thought to be important in hot ionized gas such as YSMCs. Let $Y_{\rm sp}$ be the average sputtering yield by impinging H with mean thermal velocity $\langle v\rangle$. Then, the sputtering rate is given by
\begin{eqnarray}
\frac{4\pi \rho a^{2}da}{dt} = n_{\rm H}\langle v\rangle\pi a^{2}Y_{\rm sp}m_{\rm H},
\end{eqnarray}
which leads to the sputtering time:
\begin{eqnarray}
\tau_{\rm sp}&=&\frac{a}{da/dt}\nonumber\\
&\simeq& 9.8\times 10^{3}a_{-5}n_{1}^{-1}\left(\frac{10^{6}~\rm K}{T_{\rm gas}}\right)^{1/2}\left(\frac{0.1}{Y_{\rm sp}}\right) \rm yr,~~~\label{eq:tausp}
\end{eqnarray}
where $n_{1}=n_{\rm H}/(1~\rm cm^{-3})$, and $Y_{\rm sp}$ falls rapidly after its peak value of $\sim 0.1$ at the Bohr velocity due to the decrease of the nucleons-nucleon cross-section\cite{2015ApJ...806..255H}.

In the case of SNe, grains can be destroyed by grain-grain collisions and non-thermal sputtering due to drift of accelerated grains through the gas\cite{Hoang:2017kg}. The destruction time by grain-grain collisions can be estimated as the mean time between two collisions:
\begin{eqnarray}
\tau_{\rm gg}=\frac{1}{\pi a^{2} n_{\rm gr}v_{\rm drift}}=\frac{4\rho a M_{g/d}}{3n_{\rm H}m_{\rm H}v_{\rm drift}}\simeq 7.6\times 10^{4}\hat{\rho}a_{-5}n_{1}^{-1}v_{\rm drift,3}^{-1}{~\rm yr},
\end{eqnarray}
where $n_{\rm gr}$ is the number density of dust grains, $M_{g/d}=100$ is the gas-to-dust mass ratio, and we have assumed the single-grain size distribution.

Grains initially located at distance of $r_{i}=1~\rm pc$ can be accelerated to $v\sim 1700~\rm km~s^{-1}$ by supernova radiation pressure\cite{Hoang:2017kg}. Thus, grain-grain collisions require a much longer timescale to form small grains compared to RATD (see equation (\ref{eq:tdisr})). In order for grain-grain collisions to reproduce small grains as required by early phase observations toward SNe Ia, i.e., $\tau_{\rm gg}< 100$ days, one needs $n_{\rm H}> 10^{8}~\rm cm^{-3}$ which corresponds to extremely dense clouds.

The destruction time by non-thermal sputtering is equal to:
\begin{eqnarray}
t_{\rm nontherm-sp}&=& \frac{\rho a}{n_{\rm H}v_{\rm drift}m_{\rm H}Y_{\rm sp}}\\
&\simeq& 5.7\times 10^{3}\hat{\rho}\left(\frac{a_{-5}}{n_{1}}\right)\left(\frac{0.1}{Y_{\rm sp}}\right)\left(\frac{10^{3}\rm km~s^{-1}}{v_{\rm drift}}\right) \rm yr.\nonumber
\end{eqnarray}

Finally, we note again that a popular mechanism of grain shattering due to supernova shocks is expected to work at a much later stage when the supernova ejecta already reaches the interstellar medium. This is different from our RATD mechanism that works within a few weeks after the explosion.

\subsection{Effect of RATD on Ly$\alpha$ photon escape from star-forming galaxies}

Hydrogen Ly$\alpha$ line ($\lambda=1216$~\AA) in the spectra of star-forming galaxies, either nearby or high-$z$, carries a wealth of information about both photon sources and surrounding media. Since Ly$\alpha$ photons experience a large number of local scatterings before escape in these objects, the emergent line profile and flux are heavily dependent upon local properties such as kinematics, column densities of neutral hydrogen, clumpy structures, and dust properties, more than the global properties of the ISM in the host galaxy. The physical quantities such as Ly$\alpha$ escape fraction ($f_{\rm esc}(\rm Ly\alpha)$) and the star-formation rate are usually inferred from the dust-corrected rest-frame UV/IR flux and/or from the dust-corrected H$\alpha$ emission flux. The dust-correction depends the UV dust extinction, which is dependent upon the grain size distribution, especially the population of small and small grains that dominates the UV extinction.

The Ly$\alpha$-selected and H$\alpha$-selected galaxies show disjoint quantities of extinction, Ly$\alpha$ escape fraction, and the star-formation rate\cite{2010Natur.464..562H}. In addition, the Ly$\alpha$-selected sample is significantly less dusty and exhibit higher escape fraction than the H$\alpha$-selected sample due to the fact that dust extinction is larger at shorter wavelength. It is well-known that the escape fraction is anti-correlated with the dust reddening\cite{2010Natur.464..562H,Atek:2014ga}. Moreover, a detailed study \cite{2017ApJ...844..171Y} reveals that the low value of $f_{\rm esc}(\rm Ly\alpha)$ can be better fit with the SMC-like extinction curve, and many galaxies have $f_{\rm esc}$ even below the fit by SMC-like extinction (see their Figure 7a). This demonstrates the greater enhancement of small grains in these star-forming galaxies. Furthermore, direct measurements of dust properties in high-z Ly$\alpha$ break galaxies indicate the typical grain size $a\sim 0.05~\mu$m\cite{2014MNRAS.439.3073Y}, which is much smaller than the ISM dust. Due to the dominance of massive stars and SNe in the star-forming galaxies, either nearby or high-z, the efficient RATD mechanism is expected to be important in disrupting large grains to produce such small grains. 

Another issue is the observation of the singly-peaked Ly$\alpha$ emission lines observed in these starburst galaxies. To explain this feature, usually a central cavity of dust grains interior of a supershell/superbubble is suggested to efficiently suppress the secondary, tertiary, and even higher order peaks redshifted with respect to the rest-frame Ly$\alpha$ line center\cite{2004ApJ...601L..25A}. The size and the age of the supershell/superbubble was roughly estimated to be $1.8~\rm pc<R<180~\rm pc$ and $10^{2}~{\rm yr}<t<5\times 10^{4}~{\rm yr}$, respectively, for a range of input energy as a free parameter from stellar radiation and explosions\cite{2000ApJ...530L...9A}. A young massive star cluster has $L\sim 10^{6}-10^{9}L_\odot$. Thus, for an intense radiation field of $L= 10^{9}L_{\odot}$, RATD can destroy large grains to form very small grains of $a<0.1216/2\pi \sim 0.02~ \mu$m. This reduces the absorption of Ly$\alpha$ photons, resulting in a cavity size of $1$ pc (see Figure \ref{fig:dust_HII_SNe} for compact grains). The central cavity can be larger if grains are composite. 

\subsection{Data availability} The data that support the plots within this paper and other findings of this study are available from the corresponding author upon reasonable request.

\begin{addendum}
\item[Correspondence] Correspondence and requests for materials should be addressed to T.H. via thiemhoang@kasi.re.kr.

\item[ Acknowledgments]
We are grateful to the reviewers for their thorough readings and constructive comments that helped us improve the paper. We thank B-G Andersson, M. Bulla, B. Burkhart, B.T. Draine, A. Goobar, V. Guillet, A. Lazarian, P. Lesaffre, R. Smith, and W. Zheng for useful comments and discussions. This work was supported by the Basic Science Research Program through the National Research Foundation of Korea (NRF), funded by the Ministry of Education (2017R1D1A1B03035359).

\item[Author contributions]
All authors contributed to the work presented in this paper. T.H. formulated the problem, carried out analytical calculations, and led the writing of the manuscript. Tr.L.N. carried out numerical calculations for SNe and KNe. H.S.L. carried out calculations for massive stars. S.H.A. contributed to the discussion of the RATD mechanism on Ly$\alpha$ photon escape. 

\item[Competing Interests] The authors declare no
competing financial interests.
 
\end{addendum}
\section*{\large References}

\newpage
\begin{table*}
\begin{center}
\caption{Grain disruption size and disruption time for dust clouds at different distances from the radiation source}\label{tab:adisr_ISM}
\begin{tabular}{l l l l l l l l l l l l l l l} \hline\hline\\
\multicolumn{3}{l}{Sources} & 
\multicolumn{3}{l}{$a_{\rm disr}(\mu\rm m)$} & \multicolumn{3}{r}{$t_{\rm disr}(\rm days)^{a}$}\\
d(pc) & 0.1 & 0.5 & 1.0 & 3.0 & 5.0 & & 0.1 & 0.5 & 1.0 & 3.0 & 5.0\cr
\hline\\
Case 1$^b$ & & & & & & & & \cr
YMSC$^c$ & 0.013 & 0.019 & 0.022 & 0.029 & 0.033 & & 114 & 2864 & 11457 & 103109 & 286416\cr
SNe Ia & 0.014 & 0.021 & 0.083 & ND & ND & & 5.15 & 17.3 & 64.03 & ND & ND\cr
SNe II-P & 0.014 & 0.025 & 0.10& ND & ND & & 1.14 & 30.52 & 198.3 & ND & ND\cr
SLSN &  0.01 & 0.014 & 0.017 & 0.048 & ND& & 0.07& 1.78 & 7.22 & 75.87 & ND\cr
\cr
Case 2$^d$ & & & & & & & & \cr
YMSC & 0.005 & 0.008 & 0.009 & 0.012 & 0.014 & & 11 & 286 & 1146 & 10311 & 28641\cr
SNe Ia & 0.007 & 0.009 & 0.011 & 0.07 & ND & & 2.37 & 7.05 & 11.69 & 48.42 & ND\cr
SNe II-P & 0.006 & 0.009 & 0.012 & 0.085 & ND & & 0.11 & 2.85 & 11.64 & 127.31 & ND\cr
SLSN & 0.003& 0.007 & 0.008 & 0.01 & 0.011 & & 0.0071 & 0.18 & 0.71 & 6.47 & 18.46\cr

\cr
\hline
\multicolumn{5}{l}{ND: No Disruption}\cr
\multicolumn{5}{l}{$^a$~Estimated for $a=0.1~\mu\rm m$}\cr
\multicolumn{5}{l}{$^b$~$S_{\rm max}=10^{9}~\rm erg~cm^{-3}$ (compact grains)}\cr
\multicolumn{5}{l}{$^c$~Results for luminosity $L=10^{7}L_{\odot}$}\cr
\multicolumn{5}{l}{$^d$~$S_{\rm max}=10^{7}~\rm erg~cm^{-3}$ (composite grains)}\cr
\cr
\hline\hline
\end{tabular}
\end{center}
\end{table*}

\begin{table}
\begin{center}
\caption{Characteristic timescales of dust destruction by different mechanisms}\label{tab:destr}
\begin{tabular}{l l } \hline\hline\\
{Mechanisms} & {{Timescales} (yr)}\cr
\hline\\
RATD & $1.0 a_{-5}^{-0.7}\bar{\lambda}_{0.5}^{1.7}U_{6}^{-1}S_{\rm max,9}^{1/2}$\cr
Thermal sputtering & $9.8\times 10^{3}a_{-5}n_{1}^{-1}T_{6}^{-1/2}(0.1/Y_{\rm sp})$ \cr
Non-thermal sputtering & $5.7\times 10^{3}\hat{\rho}a_{-5}n_{1}^{-1}v_{\rm drift,3}^{-1}(0.1/Y_{\rm sp}) $ \cr
Grain-grain collision & $7.6\times 10^{4}\hat{\rho}a_{-5}n_{1}^{-1}v_{\rm drift,3}^{-1}$\cr
\cr
\hline
\multicolumn{2}{l}{{\it Notes}:~$a_{-5}=a/(10^{-5}~\rm cm), U_{6}=U/10^{6}$}\cr
\multicolumn{2}{l}{$S_{\rm max,9}=S_{\rm max}/(10^{9}~\rm erg~cm^{-3})$} \cr
\multicolumn{2}{l}{$n_{1}=n_{\rm H}/(10~\rm cm^{-3}),T_{6}=T_{\rm gas}/(10^{6}~\rm K) $}  \cr
\multicolumn{2}{l}{$v_{\rm drift,3}=v_{\rm drift}/(10^{3}~\rm km~s^{-1})$,~and $Y_{\rm sp}$ sputtering yield} \cr\cr
\hline\hline
\end{tabular}
\end{center}
\end{table}


\newpage

\begin{thebibliography}{10}
\expandafter\ifx\csname url\endcsname\relax
  \def\url#1{\texttt{#1}}\fi
\expandafter\ifx\csname urlprefix\endcsname\relax\def\urlprefix{URL }\fi
\providecommand{\bibinfo}[2]{#2}
\providecommand{\eprint}[2][]{\url{#2}}

\bibitem{2008AJ....136.1415R}
\bibinfo{author}{Reines, A.~E.}, \bibinfo{author}{Johnson, K.~E.} \&
  \bibinfo{author}{Hunt, L.~K.}
\newblock \bibinfo{title}{{a New View of the Super Star Clusters in the
  Low-Metallicity Galaxy SBS 0335-052}}.
\newblock \emph{\bibinfo{journal}{\apj}} \textbf{\bibinfo{volume}{136}},
  \bibinfo{pages}{1415--1426} (\bibinfo{year}{2008}).

\bibitem{2013A&A...552A.140R}
\bibinfo{author}{Rela{\~n}o, M.} \emph{et~al.}
\newblock \bibinfo{title}{{Spectral energy distributions of H ii regions in M
  33 (HerM33es)}}.
\newblock \emph{\bibinfo{journal}{\aa}} \textbf{\bibinfo{volume}{552}},
  \bibinfo{pages}{A140} (\bibinfo{year}{2013}).

\bibitem{2017ApJ...843...95M}
\bibinfo{author}{Mart{\'\i}nez-Gonz{\'a}lez, S.}, \bibinfo{author}{W{\"u}nsch,
  R.} \& \bibinfo{author}{Palou{\v s}, J.}
\newblock \bibinfo{title}{{Can Dust Injected by SNe Explain the NIR-MIR Excess
  in Young Massive Stellar Clusters?}}
\newblock \emph{\bibinfo{journal}{\apj}} \textbf{\bibinfo{volume}{843}},
  \bibinfo{pages}{95} (\bibinfo{year}{2017}).

\bibitem{1979ApJ...231...77D}
\bibinfo{author}{Draine, B.~T.} \& \bibinfo{author}{Salpeter, E.~E.}
\newblock \bibinfo{title}{{On the physics of dust grains in hot gas}}.
\newblock \emph{\bibinfo{journal}{\apj}} \textbf{\bibinfo{volume}{231}},
  \bibinfo{pages}{77--94} (\bibinfo{year}{1979}).

\bibitem{1998AJ....116.1009R}
\bibinfo{author}{Riess, A.~G.}, \bibinfo{author}{Filippenko, A.~V.},
  \bibinfo{author}{Challis, P.} \& \bibinfo{author}{et~al.}
\newblock \bibinfo{title}{{Observational Evidence from Supernovae for an
  Accelerating Universe and a Cosmological Constant}}.
\newblock \emph{\bibinfo{journal}{\aj}} \textbf{\bibinfo{volume}{116}},
  \bibinfo{pages}{1009--1038} (\bibinfo{year}{1998}).

\bibitem{2008A&A...487...19N}
\bibinfo{author}{Nobili, S.} \& \bibinfo{author}{Goobar, A.}
\newblock \bibinfo{title}{{The colour-lightcurve shape relation of type Ia
  supernovae and the reddening law}}.
\newblock \emph{\bibinfo{journal}{\aa}} \textbf{\bibinfo{volume}{487}},
  \bibinfo{pages}{19--31} (\bibinfo{year}{2008}).

\bibitem{2014ApJ...789...32B}
\bibinfo{author}{Burns, C.~R.} \emph{et~al.}
\newblock \bibinfo{title}{{The Carnegie Supernova Project: Intrinsic Colors of
  Type Ia Supernovae}}.
\newblock \emph{\bibinfo{journal}{\apj}} \textbf{\bibinfo{volume}{789}},
  \bibinfo{pages}{32} (\bibinfo{year}{2014}).

\bibitem{2003ARA&A..41..241D}
\bibinfo{author}{Draine, B.~T.}
\newblock \bibinfo{title}{{Interstellar Dust Grains}}.
\newblock \emph{\bibinfo{journal}{\araa}} \textbf{\bibinfo{volume}{41}},
  \bibinfo{pages}{241--289} (\bibinfo{year}{2003}).

\bibitem{Kawabata:2014gy}
\bibinfo{author}{Kawabata, K.~S.} \emph{et~al.}
\newblock \bibinfo{title}{{OPTICAL AND NEAR-INFRARED POLARIMETRY OF HIGHLY
  REDDENED Type Ia SUPERNOVA 2014J: PECULIAR PROPERTIES OF DUST IN M82}}.
\newblock \emph{\bibinfo{journal}{\apj}} \textbf{\bibinfo{volume}{795}},
  \bibinfo{pages}{L4} (\bibinfo{year}{2014}).

\bibitem{Patat:2015bb}
\bibinfo{author}{Patat, F.} \emph{et~al.}
\newblock \bibinfo{title}{{Properties of extragalactic dust inferred from
  linear polarimetry of Type Ia Supernovae}}.
\newblock \emph{\bibinfo{journal}{\aa}} \textbf{\bibinfo{volume}{577}},
  \bibinfo{pages}{A53} (\bibinfo{year}{2015}).

\bibitem{2016P&SS..133...36N}
\bibinfo{author}{Nozawa, T.}
\newblock \bibinfo{title}{{Properties of interstellar dust responsible for
  extinction laws with unusually low total-to-selective extinction ratios of
  RV=1-2}}.
\newblock \emph{\bibinfo{journal}{Planetary and Space Science}}
  \textbf{\bibinfo{volume}{133}}, \bibinfo{pages}{36--46}
  (\bibinfo{year}{2016}).

\bibitem{2017ApJ...836...13H}
\bibinfo{author}{Hoang, T.}
\newblock \bibinfo{title}{{Properties and Alignment of Interstellar Dust Grains
  toward Type Ia Supernovae with Anomalous Polarization Curves}}.
\newblock \emph{\bibinfo{journal}{\apj}} \textbf{\bibinfo{volume}{836}},
  \bibinfo{pages}{13} (\bibinfo{year}{2017}).

\bibitem{2010A&A...518L.138B}
\bibinfo{author}{Barlow, M.~J.} \emph{et~al.}
\newblock \bibinfo{title}{{A Herschel PACS and SPIRE study of the dust content
  of the Cassiopeia A supernova remnant}}.
\newblock \emph{\bibinfo{journal}{\aa}} \textbf{\bibinfo{volume}{518}},
  \bibinfo{pages}{L138} (\bibinfo{year}{2010}).

\bibitem{2011Sci...333.1258M}
\bibinfo{author}{Matsuura, M.} \emph{et~al.}
\newblock \bibinfo{title}{{Herschel Detects a Massive Dust Reservoir in
  Supernova 1987A}}.
\newblock \emph{\bibinfo{journal}{Science}} \textbf{\bibinfo{volume}{333}},
  \bibinfo{pages}{1258--} (\bibinfo{year}{2011}).

\bibitem{2012ApJ...760...96G}
\bibinfo{author}{Gomez, H.~L.} \emph{et~al.}
\newblock \bibinfo{title}{{A Cool Dust Factory in the Crab Nebula: A Herschel
  Study of the Filaments}}.
\newblock \emph{\bibinfo{journal}{\apj}} \textbf{\bibinfo{volume}{760}},
  \bibinfo{pages}{96} (\bibinfo{year}{2012}).

\bibitem{Chawner:2018dn}
\bibinfo{author}{Chawner, H.} \emph{et~al.}
\newblock \bibinfo{title}{{A catalogue of Galactic supernova remnants in the
  far-infrared: revealing ejecta dust in pulsar wind nebulae}}.
\newblock \emph{\bibinfo{journal}{\mnras}} \textbf{\bibinfo{volume}{483}},
  \bibinfo{pages}{70--118} (\bibinfo{year}{2018}).

\bibitem{2003MNRAS.343..427M}
\bibinfo{author}{Morgan, H.~L.} \& \bibinfo{author}{Edmunds, M.~G.}
\newblock \bibinfo{title}{{Dust formation in early galaxies}}.
\newblock \emph{\bibinfo{journal}{\mnras}} \textbf{\bibinfo{volume}{343}},
  \bibinfo{pages}{427--442} (\bibinfo{year}{2003}).

\bibitem{2014MNRAS.439.3073Y}
\bibinfo{author}{Yajima, H.}, \bibinfo{author}{Nagamine, K.},
  \bibinfo{author}{Thompson, R.} \& \bibinfo{author}{Choi, J.-H.}
\newblock \bibinfo{title}{{Dust properties of Lyman-break galaxies in
  cosmological simulations}}.
\newblock \emph{\bibinfo{journal}{\mnras}} \textbf{\bibinfo{volume}{439}},
  \bibinfo{pages}{3073--3084} (\bibinfo{year}{2014}).

\bibitem{2009ApJ...694.1067P}
\bibinfo{author}{Poznanski, D.} \emph{et~al.}
\newblock \bibinfo{title}{{Improved Standardization of Type II-P Supernovae:
  Application to an Expanded Sample}}.
\newblock \emph{\bibinfo{journal}{\apj}} \textbf{\bibinfo{volume}{694}},
  \bibinfo{pages}{1067--1079} (\bibinfo{year}{2009}).

\bibitem{2010ApJ...715..833O}
\bibinfo{author}{Olivares~E, F.} \emph{et~al.}
\newblock \bibinfo{title}{{The Standardized Candle Method for Type II Plateau
  Supernovae}}.
\newblock \emph{\bibinfo{journal}{\apj}} \textbf{\bibinfo{volume}{715}},
  \bibinfo{pages}{833--853} (\bibinfo{year}{2010}).

\bibitem{2013ApJ...774....8T}
\bibinfo{author}{Temim, T.} \& \bibinfo{author}{Dwek, E.}
\newblock \bibinfo{title}{{The Importance of Physical Models for Deriving Dust
  Masses and Grain Size Distributions in Supernova Ejecta. I. Radiatively
  Heated Dust in the Crab Nebula}}.
\newblock \emph{\bibinfo{journal}{\apj}} \textbf{\bibinfo{volume}{774}},
  \bibinfo{pages}{8} (\bibinfo{year}{2013}).

\bibitem{2015ApJ...801..141O}
\bibinfo{author}{Owen, P.~J.} \& \bibinfo{author}{Barlow, M.~J.}
\newblock \bibinfo{title}{{The Dust and Gas Content of the Crab Nebula}}.
\newblock \emph{\bibinfo{journal}{\apj}} \textbf{\bibinfo{volume}{801}},
  \bibinfo{pages}{141} (\bibinfo{year}{2015}).

\bibitem{Gall:2014dk}
\bibinfo{author}{Gall, C.} \emph{et~al.}
\newblock \bibinfo{title}{{Rapid formation of large dust grains in the luminous
  supernova 2010jl}}.
\newblock \emph{\bibinfo{journal}{Nature}} \bibinfo{pages}{1--16}
  (\bibinfo{year}{2014}).

\bibitem{1979ApJ...231..438D}
\bibinfo{author}{Draine, B.~T.} \& \bibinfo{author}{Salpeter, E.~E.}
\newblock \bibinfo{title}{{Destruction mechanisms for interstellar dust}}.
\newblock \emph{\bibinfo{journal}{\apj}} \textbf{\bibinfo{volume}{231}},
  \bibinfo{pages}{438--455} (\bibinfo{year}{1979}).

\bibitem{1989ApJ...345..230G}
\bibinfo{author}{Guhathakurta, P.} \& \bibinfo{author}{Draine, B.~T.}
\newblock \bibinfo{title}{{Temperature fluctuations in interstellar grains. I -
  Computational method and sublimation of small grains}}.
\newblock \emph{\bibinfo{journal}{\apj}} \textbf{\bibinfo{volume}{345}},
  \bibinfo{pages}{230--244} (\bibinfo{year}{1989}).

\bibitem{2015ApJ...806..255H}
\bibinfo{author}{Hoang, T.}, \bibinfo{author}{Lazarian, A.} \&
  \bibinfo{author}{Schlickeiser, R.}
\newblock \bibinfo{title}{{On Origin and Destruction of Relativistic Dust and
  its Implication for Ultrahigh Energy Cosmic Rays}}.
\newblock \emph{\bibinfo{journal}{\apj}} \textbf{\bibinfo{volume}{806}},
  \bibinfo{pages}{255} (\bibinfo{year}{2015}).

\bibitem{1998ApJ...507L..59L}
\bibinfo{author}{Li, L.-X.} \& \bibinfo{author}{Paczy{\'n}ski, B.}
\newblock \bibinfo{title}{{Transient Events from Neutron Star Mergers}}.
\newblock \emph{\bibinfo{journal}{\apj}} \textbf{\bibinfo{volume}{507}},
  \bibinfo{pages}{L59--L62} (\bibinfo{year}{1998}).

\bibitem{2017PASJ...69..102T}
\bibinfo{author}{Tanaka, M.} \emph{et~al.}
\newblock \bibinfo{title}{{Kilonova from post-merger ejecta as an optical and
  near-Infrared counterpart of GW170817}}.
\newblock \emph{\bibinfo{journal}{\pasj}} \textbf{\bibinfo{volume}{69}},
  \bibinfo{pages}{102} (\bibinfo{year}{2017}).

\bibitem{2017ApJ...849L..19G}
\bibinfo{author}{Gall, C.}, \bibinfo{author}{Hjorth, J.},
  \bibinfo{author}{Rosswog, S.}, \bibinfo{author}{Tanvir, N.~R.} \&
  \bibinfo{author}{Levan, A.~J.}
\newblock \bibinfo{title}{{Lanthanides or Dust in Kilonovae: Lessons Learned
  from GW170817}}.
\newblock \emph{\bibinfo{journal}{\apjl}} \textbf{\bibinfo{volume}{849}},
  \bibinfo{pages}{L19} (\bibinfo{year}{2017}).

\bibitem{1976Ap&SS..43..291D}
\bibinfo{author}{Dolginov, A.~Z.} \& \bibinfo{author}{Mitrofanov, I.~G.}
\newblock \bibinfo{title}{{Orientation of cosmic dust grains}}.
\newblock \emph{\bibinfo{journal}{Ap\&SS}} \textbf{\bibinfo{volume}{43}},
  \bibinfo{pages}{291--317} (\bibinfo{year}{1976}).

\bibitem{1996ApJ...470..551D}
\bibinfo{author}{Draine, B.~T.} \& \bibinfo{author}{Weingartner, J.~C.}
\newblock \bibinfo{title}{{Radiative Torques on Interstellar Grains. I.
  Superthermal Spin-up}}.
\newblock \emph{\bibinfo{journal}{\apj}} \textbf{\bibinfo{volume}{470}},
  \bibinfo{pages}{551} (\bibinfo{year}{1996}).

\bibitem{2007MNRAS.378..910L}
\bibinfo{author}{Lazarian, A.} \& \bibinfo{author}{Hoang, T.}
\newblock \bibinfo{title}{{Radiative torques: analytical model and basic
  properties}}.
\newblock \emph{\bibinfo{journal}{\mnras}} \textbf{\bibinfo{volume}{378}},
  \bibinfo{pages}{910--946} (\bibinfo{year}{2007}).

\bibitem{1983A&A...128..212M}
\bibinfo{author}{Mathis, J.~S.}, \bibinfo{author}{Mezger, P.~G.} \&
  \bibinfo{author}{Panagia, N.}
\newblock \bibinfo{title}{{Interstellar radiation field and dust temperatures
  in the diffuse interstellar matter and in giant molecular clouds}}.
\newblock \emph{\bibinfo{journal}{\aa}} \textbf{\bibinfo{volume}{128}},
  \bibinfo{pages}{212--229} (\bibinfo{year}{1983}).

\bibitem{2009ApJ...695.1457H}
\bibinfo{author}{Hoang, T.} \& \bibinfo{author}{Lazarian, A.}
\newblock \bibinfo{title}{{Radiative Torques Alignment in the Presence of
  Pinwheel Torques}}.
\newblock \emph{\bibinfo{journal}{\apj}} \textbf{\bibinfo{volume}{695}},
  \bibinfo{pages}{1457--1476} (\bibinfo{year}{2009}).

\bibitem{2014ApJ...784..147S}
\bibinfo{author}{Stephens, I.~W.} \emph{et~al.}
\newblock \bibinfo{title}{{Spitzer Observations of Dust Emission from H II
  Regions in the Large Magellanic Cloud}}.
\newblock \emph{\bibinfo{journal}{\apj}} \textbf{\bibinfo{volume}{784}},
  \bibinfo{pages}{147} (\bibinfo{year}{2014}).

\bibitem{2007ApJ...665..390L}
\bibinfo{author}{Lebouteiller, V.}, \bibinfo{author}{Brandl, B.},
  \bibinfo{author}{Bernard-Salas, J.}, \bibinfo{author}{Devost, D.} \&
  \bibinfo{author}{Houck, J.~R.}
\newblock \bibinfo{title}{{PAH Strength and the Interstellar Radiation Field
  around the Massive Young Cluster NGC 3603}}.
\newblock \emph{\bibinfo{journal}{\apj}} \textbf{\bibinfo{volume}{665}},
  \bibinfo{pages}{390--401} (\bibinfo{year}{2007}).

\bibitem{Relano:2018kx}
\bibinfo{author}{Rela{\~n}o, M.} \emph{et~al.}
\newblock \bibinfo{title}{{Spatially resolving the dust properties and
  submillimetre excess in M 33}}.
\newblock \emph{\bibinfo{journal}{arXiv:1801.04806}}  (\bibinfo{year}{2018}).
\newblock \eprint{1801.04806v1}.

\bibitem{2010AJ....139..120F}
\bibinfo{author}{Folatelli, G.} \emph{et~al.}
\newblock \bibinfo{title}{{The Carnegie Supernova Project: Analysis of the
  First Sample of Low-Redshift Type-Ia Supernovae}}.
\newblock \emph{\bibinfo{journal}{\aj}} \textbf{\bibinfo{volume}{139}},
  \bibinfo{pages}{120--144} (\bibinfo{year}{2010}).

\bibitem{2013ApJ...779...38P}
\bibinfo{author}{Phillips, M.~M.} \emph{et~al.}
\newblock \bibinfo{title}{{On the Source of the Dust Extinction in Type Ia
  Supernovae and the Discovery of Anomalously Strong Na I Absorption}}.
\newblock \emph{\bibinfo{journal}{\apj}} \textbf{\bibinfo{volume}{779}},
  \bibinfo{pages}{38} (\bibinfo{year}{2013}).

\bibitem{2015ApJ...807L..26G}
\bibinfo{author}{Gao, J.}, \bibinfo{author}{Jiang, B.~W.}, \bibinfo{author}{Li,
  A.}, \bibinfo{author}{Li, J.} \& \bibinfo{author}{Wang, X.}
\newblock \bibinfo{title}{{Physical Dust Models for the Extinction toward
  Supernova 2014J in M82}}.
\newblock \emph{\bibinfo{journal}{\apjl}} \textbf{\bibinfo{volume}{807}},
  \bibinfo{pages}{L26} (\bibinfo{year}{2015}).

\bibitem{2008ApJ...675..626W}
\bibinfo{author}{Wang, X.} \emph{et~al.}
\newblock \bibinfo{title}{{Optical and Near-Infrared Observations of the Highly
  Reddened, Rapidly Expanding Type Ia Supernova SN 2006X in M100}}.
\newblock \emph{\bibinfo{journal}{\apj}} \textbf{\bibinfo{volume}{675}},
  \bibinfo{pages}{626--643} (\bibinfo{year}{2008}).

\bibitem{2018MNRAS.tmp.1547B}
\bibinfo{author}{Bulla, M.}, \bibinfo{author}{Goobar, A.} \&
  \bibinfo{author}{Dhawan, S.}
\newblock \bibinfo{title}{{Shedding light on the Type Ia supernova extinction
  puzzle: dust location found}}.
\newblock \emph{\bibinfo{journal}{\mnras}} \textbf{\bibinfo{volume}{479}},
  \bibinfo{pages}{3663--3674} (\bibinfo{year}{2018}).

\bibitem{2012ApJ...746...48M}
\bibinfo{author}{Metzger, B.~D.} \& \bibinfo{author}{Berger, E.}
\newblock \bibinfo{title}{{What is the Most Promising Electromagnetic
  Counterpart of a Neutron Star Binary Merger?}}
\newblock \emph{\bibinfo{journal}{\apj}} \textbf{\bibinfo{volume}{746}},
  \bibinfo{pages}{48} (\bibinfo{year}{2012}).

\bibitem{Levan:2017jz}
\bibinfo{author}{Levan, A.~J.} \emph{et~al.}
\newblock \bibinfo{title}{{The Environment of the Binary Neutron Star Merger
  GW170817}}.
\newblock \emph{\bibinfo{journal}{\apjl}} \textbf{\bibinfo{volume}{848}},
  \bibinfo{pages}{0--0} (\bibinfo{year}{2017}).

\bibitem{Covino:2017dq}
\bibinfo{author}{Covino, S.} \emph{et~al.}
\newblock \bibinfo{title}{{The unpolarized macronova associated with the
  gravitational wave event GW 170817}}.
\newblock \emph{\bibinfo{journal}{Nature Astronomy}} \bibinfo{pages}{1--4}
  (\bibinfo{year}{2017}).

\bibitem{1997ApJ...487..625G}
\bibinfo{author}{Gordon, K.~D.}, \bibinfo{author}{Calzetti, D.} \&
  \bibinfo{author}{Witt, A.~N.}
\newblock \bibinfo{title}{{Dust in Starburst Galaxies}}.
\newblock \emph{\bibinfo{journal}{\apj}} \textbf{\bibinfo{volume}{487}},
  \bibinfo{pages}{625} (\bibinfo{year}{1997}).

\bibitem{2017ApJ...844..171Y}
\bibinfo{author}{Yang, H.} \emph{et~al.}
\newblock \bibinfo{title}{{Ly$\alpha$ Profile, Dust, and Prediction of
  Ly$\alpha$ Escape Fraction in Green Pea Galaxies}}.
\newblock \emph{\bibinfo{journal}{\apj}} \textbf{\bibinfo{volume}{844}},
  \bibinfo{pages}{171} (\bibinfo{year}{2017}).

\bibitem{2012A&A...537A..15S}
\bibinfo{author}{Schady, P.} \emph{et~al.}
\newblock \bibinfo{title}{{The dust extinction curves of gamma-ray burst host
  galaxies}}.
\newblock \emph{\bibinfo{journal}{\aa}} \textbf{\bibinfo{volume}{537}},
  \bibinfo{pages}{15} (\bibinfo{year}{2012}).

\bibitem{2004AJ....128.1112H}
\bibinfo{author}{Hopkins, P.~F.} \emph{et~al.}
\newblock \bibinfo{title}{{Dust Reddening in Sloan Digital Sky Survey
  Quasars}}.
\newblock \emph{\bibinfo{journal}{\aj}} \textbf{\bibinfo{volume}{128}},
  \bibinfo{pages}{1112--1123} (\bibinfo{year}{2004}).


\expandafter\ifx\csname url\endcsname\relax
  \def\url#1{\texttt{#1}}\fi
\expandafter\ifx\csname urlprefix\endcsname\relax\def\urlprefix{URL }\fi
\providecommand{\bibinfo}[2]{#2}
\providecommand{\eprint}[2][]{\url{#2}}
\bibitem{2018ApJ...853...56R}
\bibinfo{author}{Reddy, N.~A.} \emph{et~al.}
\newblock \bibinfo{title}{{The HDUV Survey: A Revised Assessment of the
  Relationship between UV Slope and Dust Attenuation for High-redshift
  Galaxies}}.
\newblock \emph{\bibinfo{journal}{\apj}} \textbf{\bibinfo{volume}{853}},
  \bibinfo{pages}{56} (\bibinfo{year}{2018}).

\bibitem{2004ApJ...614..781A}
\bibinfo{author}{Abbas, M.~M.} \emph{et~al.}
\newblock \bibinfo{title}{{Laboratory Experiments on Rotation and Alignment of
  the Analogs of Interstellar Dust Grains by Radiation}}.
\newblock \emph{\bibinfo{journal}{\apj}} \textbf{\bibinfo{volume}{614}},
  \bibinfo{pages}{781--795} (\bibinfo{year}{2004}).

\bibitem{Hoang:2008gb}
\bibinfo{author}{Hoang, T.} \& \bibinfo{author}{Lazarian, A.}
\newblock \bibinfo{title}{{Radiative torque alignment: essential physical
  processes}}.
\newblock \emph{\bibinfo{journal}{\mnras}} \textbf{\bibinfo{volume}{388}},
  \bibinfo{pages}{117--143} (\bibinfo{year}{2008}).

\bibitem{Andersson:2015bq}
\bibinfo{author}{Andersson, B.-G.}, \bibinfo{author}{Lazarian, A.} \&
  \bibinfo{author}{Vaillancourt, J.~E.}
\newblock \bibinfo{title}{{Interstellar Dust Grain Alignment}}.
\newblock \emph{\bibinfo{journal}{AR\aa}} \textbf{\bibinfo{volume}{53}},
  \bibinfo{pages}{501--539} (\bibinfo{year}{2015}).

\bibitem{LAH15}
\bibinfo{author}{{Lazarian}, A.}, \bibinfo{author}{{Andersson}, B.-G.} \&
  \bibinfo{author}{{Hoang}, T.}
\newblock \bibinfo{title}{Grain alignment: role of radiative torques and
  paramagnetic relaxation}.
\newblock In \bibinfo{editor}{{Kolokolova}, L.}, \bibinfo{editor}{{Hough}, J.}
  \& \bibinfo{editor}{{Levasseur-Regourd}, A.-C.} (eds.)
  \emph{\bibinfo{booktitle}{Polarimetry of stars and planetary systems}},
  \bibinfo{pages}{81} (\bibinfo{publisher}{(New York: Cambridge Univ. Press)},
  \bibinfo{year}{2015}).

\bibitem{Herranen:2018wd}
\bibinfo{author}{Herranen, J.}, \bibinfo{author}{Lazarian, A.} \&
  \bibinfo{author}{Hoang, T.}
\newblock \bibinfo{title}{{Radiative torques of irregular grains: Describing
  the alignment of a grain ensemble}}.
\newblock \emph{\bibinfo{journal}{arXiv:1812.07274}}  (\bibinfo{year}{2018}).
\newblock \eprint{1812.07274v1}.

\bibitem{2014MNRAS.438..680H}
\bibinfo{author}{Hoang, T.} \& \bibinfo{author}{Lazarian, A.}
\newblock \bibinfo{title}{{Grain alignment by radiative torques in special
  conditions and implications}}.
\newblock \emph{\bibinfo{journal}{\mnras}} \textbf{\bibinfo{volume}{438}},
  \bibinfo{pages}{680--703} (\bibinfo{year}{2014}).

\bibitem{1998ApJ...508..157D}
\bibinfo{author}{Draine, B.~T.} \& \bibinfo{author}{Lazarian, A.}
\newblock \bibinfo{title}{{Electric Dipole Radiation from Spinning Dust
  Grains}}.
\newblock \emph{\bibinfo{journal}{\apj}} \textbf{\bibinfo{volume}{508}},
  \bibinfo{pages}{157--179} (\bibinfo{year}{1998}).

\bibitem{1997ApJ...480..633D}
\bibinfo{author}{Draine, B.~T.} \& \bibinfo{author}{Weingartner, J.~C.}
\newblock \bibinfo{title}{{Radiative Torques on Interstellar Grains. II. Grain
  Alignment}}.
\newblock \emph{\bibinfo{journal}{\apj}} \textbf{\bibinfo{volume}{480}},
  \bibinfo{pages}{633} (\bibinfo{year}{1997}).

\bibitem{1979ApJ...231..404P}
\bibinfo{author}{Purcell, E.~M.}
\newblock \bibinfo{title}{{Suprathermal rotation of interstellar grains}}.
\newblock \emph{\bibinfo{journal}{\apj}} \textbf{\bibinfo{volume}{231}},
  \bibinfo{pages}{404--416} (\bibinfo{year}{1979}).

\bibitem{PortegiesZwart:2010kc}
\bibinfo{author}{Portegies~Zwart, S.~F.}, \bibinfo{author}{McMillan, S. L.~W.}
  \& \bibinfo{author}{Gieles, M.}
\newblock \bibinfo{title}{{Young Massive Star Clusters}}.
\newblock \emph{\bibinfo{journal}{\araa}} \textbf{\bibinfo{volume}{48}},
  \bibinfo{pages}{431--493} (\bibinfo{year}{2010}).

\bibitem{2009AJ....137.4517B}
\bibinfo{author}{Brown, P.~J.} \emph{et~al.}
\newblock \bibinfo{title}{{Ultraviolet Light Curves of Supernovae with the
  Swift Ultraviolet/Optical Telescope}}.
\newblock \emph{\bibinfo{journal}{\aj}} \textbf{\bibinfo{volume}{137}},
  \bibinfo{pages}{4517--4525} (\bibinfo{year}{2009}).

\bibitem{2017ApJ...848...66Z}
\bibinfo{author}{Zheng, W.}, \bibinfo{author}{Kelly, P.~L.} \&
  \bibinfo{author}{Filippenko, A.~V.}
\newblock \bibinfo{title}{{An Empirical Fitting Method for Type Ia Supernova
  Light Curves. II. Estimating the First-light Time and Rise Time}}.
\newblock \emph{\bibinfo{journal}{\apj}} \textbf{\bibinfo{volume}{848}},
  \bibinfo{pages}{66} (\bibinfo{year}{2017}).

\bibitem{1999AJ....118.2675R}
\bibinfo{author}{Riess, A.~G.} \emph{et~al.}
\newblock \bibinfo{title}{{The Rise Time of Nearby Type IA Supernovae}}.
\newblock \emph{\bibinfo{journal}{\aj}} \textbf{\bibinfo{volume}{118}},
  \bibinfo{pages}{2675--2688} (\bibinfo{year}{1999}).

\bibitem{2018MNRAS.479.2421D}
\bibinfo{author}{{Dastidar}, R.} \emph{et~al.}
\newblock \bibinfo{title}{{SN 2015ba: a Type IIP supernova with a long
  plateau}}.
\newblock \emph{\bibinfo{journal}{\mnras}} \textbf{\bibinfo{volume}{479}},
  \bibinfo{pages}{2421--2442} (\bibinfo{year}{2018}).
\newblock \eprint{1806.05470}.

\bibitem{2012Sci...337..927G}
\bibinfo{author}{Gal-Yam, A.}
\newblock \bibinfo{title}{{Luminous Supernovae}}.
\newblock \emph{\bibinfo{journal}{Science}} \textbf{\bibinfo{volume}{337}},
  \bibinfo{pages}{927--} (\bibinfo{year}{2012}).

\bibitem{2016Sci...351..257D}
\bibinfo{author}{Dong, S.} \emph{et~al.}
\newblock \bibinfo{title}{{ASASSN-15lh: A highly super-luminous supernova}}.
\newblock \emph{\bibinfo{journal}{Science}} \textbf{\bibinfo{volume}{351}},
  \bibinfo{pages}{257--260} (\bibinfo{year}{2016}).

\bibitem{2010MNRAS.406.2650M}
\bibinfo{author}{Metzger, B.~D.} \emph{et~al.}
\newblock \bibinfo{title}{{Electromagnetic counterparts of compact object
  mergers powered by the radioactive decay of r-process nuclei}}.
\newblock \emph{\bibinfo{journal}{\mnras}} \textbf{\bibinfo{volume}{406}},
  \bibinfo{pages}{2650--2662} (\bibinfo{year}{2010}).

\bibitem{2016ApJ...818..133S}
\bibinfo{author}{Silsbee, K.} \& \bibinfo{author}{Draine, B.~T.}
\newblock \bibinfo{title}{{Radiation Pressure on Fluffy Submicron-sized
  Grains}}.
\newblock \emph{\bibinfo{journal}{\apj}} \textbf{\bibinfo{volume}{818}},
  \bibinfo{pages}{133} (\bibinfo{year}{2016}).

\bibitem{Hoang:2019td}
\bibinfo{author}{Hoang, T.} \& \bibinfo{author}{Tram, L.~N.}
\newblock \bibinfo{title}{{Rotational Desorption of Ice Mantles and Complex
  Molecules from Suprathermally Rotating Dust Grains around Young Stellar
  Objects}}.
\newblock \emph{\bibinfo{journal}{arXiv.org}}  (\bibinfo{year}{2019}).
\newblock \eprint{1902.06438v1}.

\bibitem{2007ApJ...663..866D}
\bibinfo{author}{Draine, B.~T.} \emph{et~al.}
\newblock \bibinfo{title}{{Dust Masses, PAH Abundances, and Starlight
  Intensities in the SINGS Galaxy Sample}}.
\newblock \emph{\bibinfo{journal}{\apj}} \textbf{\bibinfo{volume}{663}},
  \bibinfo{pages}{866--894} (\bibinfo{year}{2007}).

\bibitem{1989ApJ...341..808M}
\bibinfo{author}{Mathis, J.~S.} \& \bibinfo{author}{Whiffen, G.}
\newblock \bibinfo{title}{{Composite interstellar grains}}.
\newblock \emph{\bibinfo{journal}{\apj}} \textbf{\bibinfo{volume}{341}},
  \bibinfo{pages}{808--822} (\bibinfo{year}{1989}).

\bibitem{Hoang:2017kg}
\bibinfo{author}{Hoang, T.}
\newblock \bibinfo{title}{{Relativistic Gas Drag on Dust Grains and
  Implications}}.
\newblock \emph{\bibinfo{journal}{\apj}} \textbf{\bibinfo{volume}{847}},
  \bibinfo{pages}{0--0} (\bibinfo{year}{2017}).

\bibitem{2000ApJ...537..796W}
\bibinfo{author}{Waxman, E.} \& \bibinfo{author}{Draine, B.~T.}
\newblock \bibinfo{title}{{Dust Sublimation by Gamma-ray Bursts and Its
  Implications}}.
\newblock \emph{\bibinfo{journal}{\apj}} \textbf{\bibinfo{volume}{537}},
  \bibinfo{pages}{796--802} (\bibinfo{year}{2000}).

\bibitem{1995ApJ...451..510S}
\bibinfo{author}{Scoville, N.} \& \bibinfo{author}{Norman, C.}
\newblock \bibinfo{title}{{Stellar Contrails in Quasi-stellar Objects: The
  Origin of Broad Absorption Lines}}.
\newblock \emph{\bibinfo{journal}{\apj}} \textbf{\bibinfo{volume}{451}},
  \bibinfo{pages}{510} (\bibinfo{year}{1995}).

\bibitem{2010Natur.464..562H}
\bibinfo{author}{Hayes, M.} \emph{et~al.}
\newblock \bibinfo{title}{{Escape of about five per cent of Lyman-$\alpha$
  photons from high-redshift star-forming galaxies}}.
\newblock \emph{\bibinfo{journal}{Nature}} \textbf{\bibinfo{volume}{464}},
  \bibinfo{pages}{562--565} (\bibinfo{year}{2010}).

\bibitem{Atek:2014ga}
\bibinfo{author}{Atek, H.} \emph{et~al.}
\newblock \bibinfo{title}{{Influence of physical galaxy properties on Ly
  $\alpha$escape in star-forming galaxies}}.
\newblock \emph{\bibinfo{journal}{\aa}} \textbf{\bibinfo{volume}{561}},
  \bibinfo{pages}{A89} (\bibinfo{year}{2014}).

\bibitem{2004ApJ...601L..25A}
\bibinfo{author}{Ahn, S.-H.}
\newblock \bibinfo{title}{{Singly Peaked Asymmetric Ly$\alpha$ from Starburst
  Galaxies}}.
\newblock \emph{\bibinfo{journal}{\apj}} \textbf{\bibinfo{volume}{601}},
  \bibinfo{pages}{L25--L28} (\bibinfo{year}{2004}).

\bibitem{2000ApJ...530L...9A}
\bibinfo{author}{Ahn, S.-H.}
\newblock \bibinfo{title}{{Environment of the Gamma-Ray Burst GRB 971214: A
  Giant H II Region Surrounded by a Galactic Supershell}}.
\newblock \emph{\bibinfo{journal}{\apj}} \textbf{\bibinfo{volume}{530}},
  \bibinfo{pages}{L9--L12} (\bibinfo{year}{2000}).

\end{thebibliography}
\end{document}


\maketitle

\begin{affiliations}
 \item 
\label{TH1}Korea Astronomy and Space Science Institute, Daejeon 34055, Republic of Korea
\item
\label{TH2} Korea University of Science and Technology, 217 Gajeong-ro, Yuseong-gu, Daejeon, 34113, Republic of Korea
\item
\label{Tr2} University of Science and Technology of Hanoi, VAST, 18 Hoang Quoc Viet, Vietnam
\item
\label{Tr3} Current address: SOFIA-USRA, NASA Ames Research Center, MS 232-11, Moffett Field, CA 94035, USA
\end{affiliations}

\newpage
\section*{Grain rotational damping}
The well-known damping process for a rotating grain is sticking collisions with gas atoms, followed by thermal evaporation\cite{Jones:1967p2924}. For a grain rotating along the z-axis with angular velocity $\omega_{z}$, the angular momentum carried away by an H atom from the grain surface is given by
\begin{eqnarray}
\delta J_{z} = I_{m}\omega_{z} = m_{\rm H}r^{2}\omega_{z}= m_{\rm H}a^{2}\rm sin^{2}\theta\omega_{z},
\end{eqnarray}
where $r$ is the distance from the atom to the spinning axis $z$, $I_{m}=m_{\rm H}r^{2}$ is the inertial moment of the hydrogen atom of mass $m_{\rm H}$, $\theta$ is the angle between the z-axis and the radius vector, and $r=asin\theta$ is the projected distance to the center. In an interval of time, there are many atoms leaving the grain surface from the different location, $\theta$. Then, assuming the isotropic distribution of $\theta$ the atoms leaving the grain, we can evaluate the mean angular momentum carried away per H atom. Thus, we can replace $\rm sin^{2}\theta = <\rm sin^{2}\theta> = 2/3$, which gives rise to
\begin{eqnarray}
\langle \delta J_{z}\rangle = \frac{2}{3}m_{\rm H}a^{2}\omega_{z}.
\end{eqnarray}

The collision rate of atomic gas with density $n_{\rm H}$ with the grain is $R_{\rm coll} = n_{\rm H}v \pi a^{2}$. Thus, one can derive the mean decrease of grain angular momentum per second as follows:
\begin{eqnarray}
\frac{\langle \Delta J_{z}\rangle}{\Delta t} = -R_{\rm coll}\langle \delta J_{z}\rangle =-\frac{2}{3}n_{\rm H}m_{\rm H}\pi a^{4}\omega_{z}\langle v\rangle.
\end{eqnarray}

The mean atomic velocity is defined by
\begin{eqnarray}
\langle v \rangle = Z \int v 4\pi v^{2}e^{-m_{\rm H}v^{2}/2kT_{\rm gas}} dv = \left(\frac{8kT_{\rm gas}}{\pi m_{\rm H}} \right)^{1/2}, 
\end{eqnarray}
where $Z=(m_{\rm H}/2\pi kT_{\rm gas})^{3/2}$ is the normalization factor of the Boltzmann distribution of gas velocity, and $T_{\rm gas}$ is the gas temperature.

Thus, for a gas with He of $10\%$ abundance, the total damping rate by gas collisions is
\begin{eqnarray}
\frac{\langle \Delta J_{z}\rangle_{\rm H+He}}{\Delta t} &=&-\frac{2}{3}1.2n_{\rm H}m_{\rm H}\pi a^{4}\omega_{z}\left(\frac{8kT_{\rm gas}}{\pi m_{\rm H}} \right)^{1/2}\\
&=&-\frac{I\omega_{z}}{\tau_{\rm gas}},
\end{eqnarray}
where the characteristic rotational damping time $\tau_{\rm gas}$ reads
\begin{eqnarray}
\tau_{\rm gas}&=&\frac{3}{4\sqrt{\pi}}\frac{I}{1.2n_{\rm H}m_{\rm H}
v_{\rm th}a^{4}}\nonumber\\
&\simeq& 8.74\times 10^{4}a_{-5}\hat{\rho}\left(\frac{30~\rm cm^{-3}}{n_{\rm H}}\right)\left(\frac{100~\rm K}{T_{\rm gas}}\right)^{1/2}~{\rm yr},~~\label{eq:tau_gas}
\end{eqnarray}
where $v_{\rm th}=\left(2k_{\rm B}T_{\rm gas}/m_{\rm H}\right)^{1/2}$ is the thermal velocity of a gas atoms\cite{2009ApJ...695.1457H,1996ApJ...470..551D}.



\section*{Tensile strengths of grain materials}\label{sec:tensile}
The tensile strength of dust grains depends on the composition and internal structures. {Ideal material without impurity}, such as diamonds, can have $S_{\rm max}\sim 10^{11}~\rm erg~cm^{-3}$. Compact grains have higher $S_{\rm max}$ than porous/composite grains, and crystalline structures have higher strength than polycrystalline ones. Note that $S_{\rm max}\sim 10^{9}-10^{10}~\rm erg~ cm^{-3}$ is suggested\cite{1974ApJ...190....1B} for polycrystalline bulk solid. {\bf Supplementary} Table \ref{tab:tensile}  lists the approximate values of the tensile strength for several materials, both monocrystalline and polycrystalline structures.

For the sake of completeness, here we describe the tensile strength of composite grains. One can estimate the tensile strength for an aggregate grain consisting of nanoparticles of average size $a_{p}$ as follows\cite{1995A&A...295L..35G}:
\begin{eqnarray}
S_{\rm max}=3\beta(1-P)\frac{\bar{E}}{2ha_{p}^{2}},
\end{eqnarray}
where $\beta$ is the mean number of contact points per nanoparticle between particles between 1-10, $\bar{E}$ is the mean intermolecular interaction energy at the contact surfaces, and $h$ is the mean intermolecular distance at the contact surface. Here $P$ is the porosity which is defined such that the mass density of the porous grain is $\rho=\rho_{0}(1-P)$ with $\rho_{0}$ being the mass density of fully compact grain. The value $P=0.2$ indicates an empty volume of $20\%$.

Assuming the interaction between constituent particles is van der Waals forces, thus $\bar{E}=\alpha 10^{-3}$ eV where $\alpha$ is the coefficient of order of unity, and $h=0.3$ nm\cite{1997A&A...323..566L}. Thus, one obtains
\begin{eqnarray}
S_{\rm max}&\simeq& 10^{7}\left(\frac{\beta}{5}\right)(1-P)\left(\frac{\alpha\bar{E}}{10^{-3}~\rm eV}\right)\nonumber\\
&&\times \left(\frac{a}{2~\rm nm}\right)^{-2}\left(\frac{0.3~\rm nm}{h}\right) ~\rm erg~cm^{-3},\label{eq:Smax_agg}
\end{eqnarray}
which yields $S_{\rm max}\sim 10^{7}~\rm erg~cm^{-3}$ for $a_{p}=2$ nm, and $S_{\rm max}\sim 1.6\times 10^{6}~\rm erg~cm^{-3}$ for $a_{p}=5$ nm.



\newpage
\begin{table}
\begin{center}
\caption{Tensile strength of different materials}\label{tab:tensile}
\begin{tabular}{l l } \hline\hline\\
{Material} & {Tensile Strength ($\rm erg~cm^{-3}$)}\cr
\hline\\
Fe (pure, single-crystal iron) & $10^{8}$ \cr
Fe (metal) & $5.4\times 10^{9}$ \cr
Fe (whisker)$^a$ & $1.3\times 10^{11}$\cr
Graphite (monocrystalline) & $10^{10}-3\times 10^{11}$\cr
Graphite (polycrystalline) & $1-5\times 10^{8}$\cr
Si (monocrystalline) $^a$ & $4.1\times 10^{10}$ \cr 
SiO$_{2}$ (glass rods and fibers)$^a$ & $1.3\times 10^{11}$ \cr
Mg$_{2}$SiO$_{4}$ (forsterite)$^b$ & $9.5\times 10^{9}$ \cr 
SiC/Mg alloy (nanoscale)$^c$ & $2.1\times 10^{9}$ \cr
Diamond & $2\times 10^{10}-6\times 10^{11}$ \cr
\cr
\hline
\multicolumn{2}{l}{$^a$~See~ ref.\cite{1972JMatS...7..239M}}\cr
\multicolumn{2}{l}{$^b$~See~ ref.\cite{Petrovic2001}}\cr
\multicolumn{2}{l}{$^c$~See~ref.\cite{Guo:2013fw}}\cr
\hline\hline
\end{tabular}
\end{center}
\end{table}

\begin{figure*}
\includegraphics[scale=0.5]{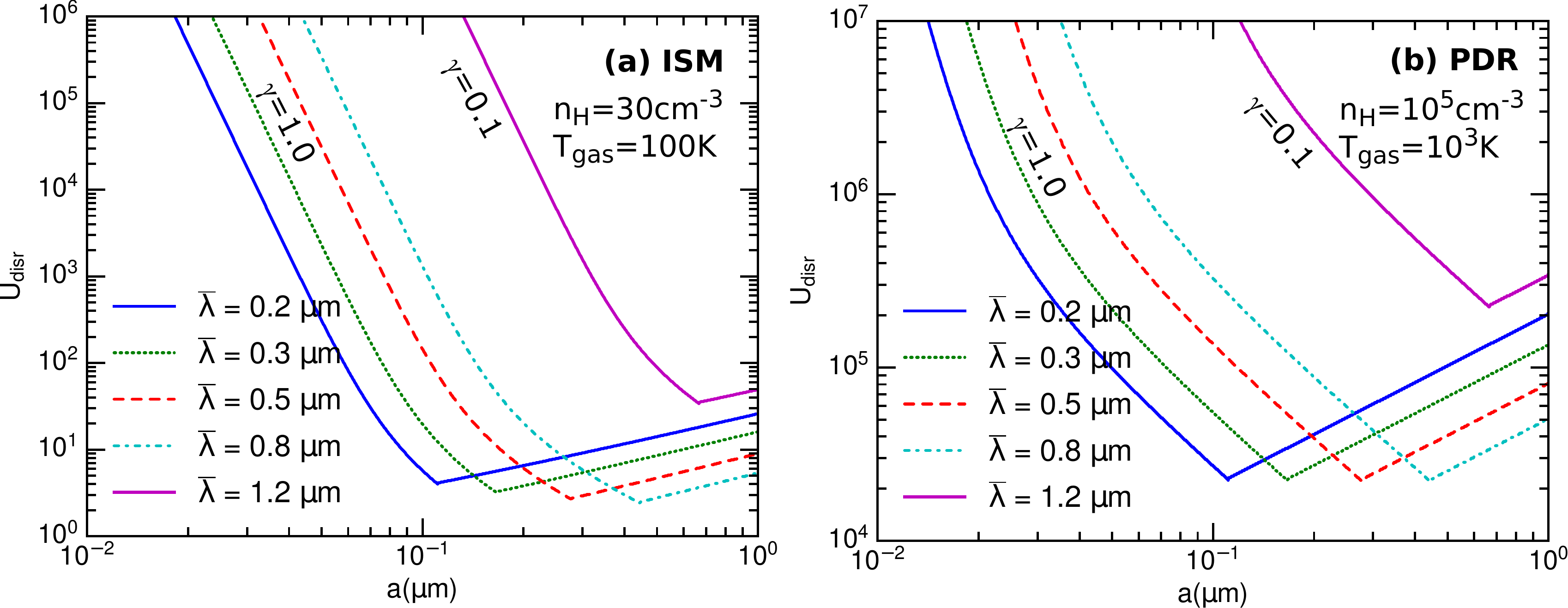}
\caption{{\bf Critical radiation strength required for disruption vs. grain size for the different radiation spectra characterized by $\bar{\lambda}=0.2-0.8~\mu$m and $\gamma=1$}. Results computed for the diffuse ISM (panel (a)) and photodissociation regions (PDRs, panel (b)). The results for the standard ISRF with $\bar{\lambda}=1.2~\mu$m and $\gamma=0.1$ are shown for comparison. The typical tensile strength $S_{\rm max}=10^{9}~\rm erg~cm^{-3}$ is assumed. The critical radiation strength for PDRs is much larger than for the diffuse ISM due to higher gas density and temperature (i.e., larger damping rate).}
\label{fig:Udisr_a}
\end{figure*}

\begin{figure}[h]
\includegraphics[scale=0.8]{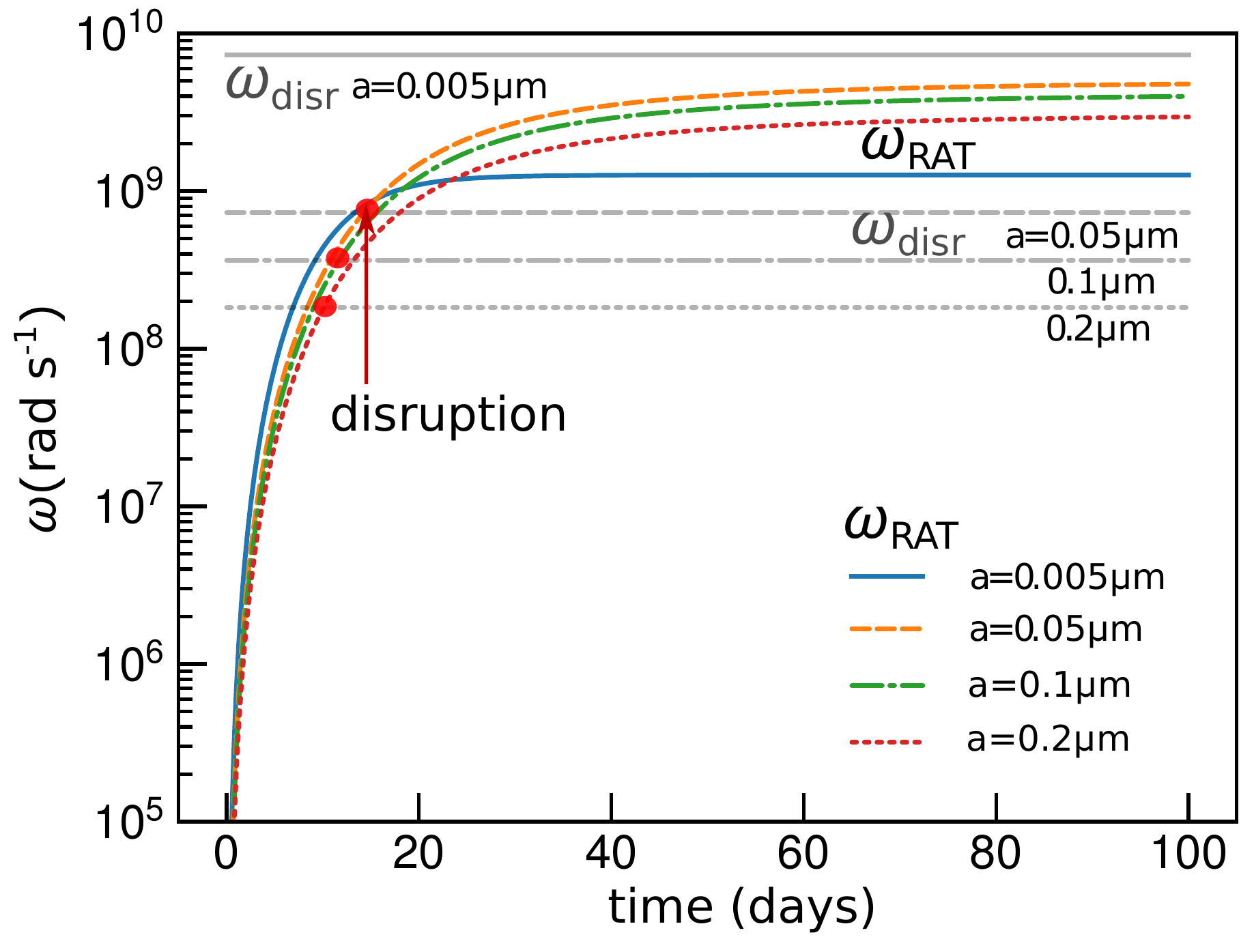}
\caption{{\bf Grain angular velocity vs. time from the explosion obtained from numerically solving equation of motion for a time-varying radiation source of SNe Ia for several grain sizes at 1 pc.} The horizontal lines show the critical velocity of disruption $\omega_{\rm disr}$ with $S_{\rm max}=10^{7}~\rm erg~cm^{-3}$, for the corresponding grain size. Grain angular velocity rapidly increases with time due to radiative torques, and filled circles mark the disruption time where $\omega_{\rm RAT}=\omega_{\rm disr}$.}
\label{fig:omega_t}
\end{figure}